%% file: gluenum_archive.tex
\documentstyle[12pt]{article}
\if@twoside
 \else
 \fi

 \parskip \medskipamount
\newcommand{\be}{\begin{eqnarray}}

\newcommand{\ee}{\end{eqnarray}}
\begin{document}


\title{\boldmath The initial gluon multiplicity in 
heavy ion collisions.}

\author{Alex Krasnitz\\
       {\small\it UCEH, Universidade do Algarve,
        Campus de Gambelas, P-8000 Faro, Portugal.}\\ 
        Raju Venugopalan\\ 
        {\small\it Physics Department,
        Brookhaven National Laboratory,
        Upton, NY 11973, USA.}\\
        }
\maketitle
\begin{abstract}

The initial gluon multiplicity per unit area per unit rapidity,
$dN/L^2/d\eta$, in high energy nuclear collisions, is equal to $f_N
(g^2\mu L) (g^2\mu)^2/g^2$, with $\mu^2$ proportional to the gluon
density per unit area of the colliding nuclei.  For an SU(2) gauge
theory, we compute $f_N (g^2\mu L)=0.14\pm 0.01$ for a wide range in
$g^2\mu L$.  Extrapolating to SU(3), we predict $dN/L^2/d\eta$ for
values of $g^2\mu L$ in the range relevant to the Relativistic Heavy
Ion Collider and the Large Hadron Collider.  We compute the initial gluon
transverse momentum distribution, $dN/L^2/d^2 k_\perp$, and show it to be 
well behaved at low $k_\perp$. 
\end{abstract}

\vspace*{0.2cm}

A topic of considerable current interest is the possibility of forming
an equilibrated plasma of quarks and gluons, a quark--gluon plasma
(QGP), in very high energy nuclear collisions. Experimental signatures
of such a plasma may provide insight into the nature of the QCD phase
diagram at finite temperature and baryon density~\cite{QM99}.

Equally interesting, is the information that heavy ion collisions may
provide about the distributions of partons in the wavefunctions of the
nuclei {\it before} the collision.  At very high energies, the growth
of parton distributions in the nuclear wavefunction saturates, forming
a state of matter sometimes 
called a Color Glass Condensate (CGC)~\cite{Larry}.  The
condensate is characterized by a bulk momentum scale $Q_s$. If $Q_s\gg
\Lambda_{QCD}$, the properties of this condensate, albeit
non--perturbative, can be studied in weak coupling. The partons that
comprise this condensate are freed in a collision.  Since the scale of
the condensate, $Q_s$, is the only scale in the problem, the initial
multiplicity and energy distributions of produced gluons at central
rapidities are determined by this scale alone. We will briefly discuss
later the relation of these quantitities to physical observables.

The above statements may be quantified in a classical effective field
theory approach (EFT) to high energy scattering~\cite{MV}. The EFT is
classical because, at central rapidities, where $x\ll 1$ and $p_\perp\gg
\Lambda_{QCD}$, ($x \sim p_\perp/\sqrt{s}$), parton distributions grow
rapidly with decreasing $x$ giving rise to large occupation numbers.
Briefly, the EFT separates partons in a hadron (or nucleus) into
static, high $x$ valence and hard glue sources, and ``wee'' small $x$
fields. For a large nucleus in the infinite momentum frame, the hard
sources with color charge density $\rho$, are randomly distributed in
the transverse plane with the distribution
\be
P([\rho])=
\exp\left(-{1\over 2g^4\mu^2}\,\int d^2 x_\perp\,\rho^2(x_\perp)\right) 
\, .
\label{Gauss}
\ee
The average squared color charge per unit area is determined by the parameter
$\mu^2$, which is the only dimensional parameter in the EFT apart from the 
linear size $L$ of the nucleus.
Parton distributions, correlation 
functions of the wee gauge fields, are computed by averaging over gauge 
fields with the weight $P$([$\rho$])~\cite{comment0}. 

Quantum corrections to the EFT~\cite{AJLR} are implemented using
Wilson renormalization group techniques~\cite{JKLMVW}.  The scale
$\mu^2$ grows with decreasing $x$, and can be estimated from the {\it
nucleon} quark and gluon distributions at $x\ll
1$~\cite{GyulassyMclerran}. The saturation scale --- at which parton
distributions stop growing rapidly with decreasing $x$ --- is $Q_s\sim 6
g^2\mu/4\pi$~\cite{MR99}, a function determined self--consistently from the 
typical $x$ and $Q^2$ of interest. Since most of the saturated partons
have momenta of this order, $Q_s$, rather than $g^2\mu$, is the
appropriate scale. At RHIC energies, $Q_s\sim 1$ GeV, and at LHC,
$Q_s\sim 2$--$3$ GeV.  The magnitude of $Q_s$ is a rough estimate
because of our uncertain knowledge of gluon structure functions at
these energies. This scale, and the properties of the CGC, may be
further determined in future high energy deeply inelastic scattering
(DIS) experiments off nuclei~\cite{HeRHIC}.

In this letter, we obtain a non-perturbative relation between the
multiplicity of gluons produced in heavy ion collisions at central
rapidities on the one hand, and $g^2\mu$, or
equivalently  $Q_s$, on the other.  We will also
demonstrate that non--perturbative strong field effects, at momenta
$k_\perp\sim Q_s$, qualitatitively alter transverse momentum distributions,
rendering them infrared finite. A preliminary version of these
results was reported in Ref.~\cite{QCD2000}.  In a previous
letter~\cite{AlexRaj3}, we obtained a similar expression for the
energy of gluons, per unit rapidity, produced shortly after a very
high energy nuclear collision.

The problem of initial conditions~\cite{BL} for nuclear scattering can be
formulated in the classical EFT~\cite{KMW} in the gauge $A^\tau=0$.
Matching the Yang--Mills equations $D_\mu F^{\mu\nu} = J^\nu$ in the
four light cone regions, along the light cone, one obtains for the
gauge fields in the forward light cone, at proper time $\tau=0$, the
relations $A^i = A_1^i + A_2^i$ and $A^{\pm}= \pm ig
x^{\pm}[A_1^i,A_2^i]/2$.  Here $J^\nu = \Sigma_{1,2}
\delta^{\nu,\pm}\delta(x^{\mp})\rho^{1,2}(x_\perp)$ are random light cone
sources corresponding to the valence or hard glue sources in the two
nuclei. The transverse pure gauge fields $A_{1,2}^i(\rho^\pm)$, with
$i=1,2$ are solutions of the Yang--Mills equations for each of the two nuclei
before the collision.  With these initial conditions, the Yang--Mills
equations can be solved in the forward light cone 
to obtain gluon configurations at late proper
times. Since the initial conditions depend on the sources $\rho^\pm$,
averages over different realizations of the sources --- specified by the
weight in Eq.~(\ref{Gauss}) --- must be performed.

Perturbative solutions for the number distributions in transverse momentum, 
per unit rapidity,  were obtained in Refs.~\cite{KMW,DYEA}. These were 
shown to be infrared divergent. In the classical EFT, this divergence is 
logarithmic. The number distributions have the form 
\be
n_{k_\perp}\propto \, {1\over \alpha_S}\, \left({\alpha_S\mu\over 
k_\perp}\right)^4\,\ln\left({k_\perp\over \alpha_S\mu}\right) \, ,
\label{LPT}
\ee
for $k_\perp\gg \alpha_S\mu$. The perturbative description breaks down at
$k_\perp\sim \alpha_S\mu$. Thus, for robust predictions of gluon multiplicity 
distributions, a fully non-perturbative study of the classical EFT 
is necessary~\cite{AlexRaj1}.

The model is discretized on a lattice in the transverse momentum
plane. Boost invariance and periodic boundary conditions are
assumed. The lattice Hamiltonian is the Kogut--Susskind Hamilitonian
in 2+1--dimensions coupled to an adjoint scalar field. The lattice
field equations are then solved by computing the Poisson brackets, with 
initial conditions that are the lattice analogs of the continuum initial 
conditions mentioned earlier. Technical details of our simulations can 
be found in Refs.~\cite{AlexRaj2,AlexRaj3}. Our simulations are presently 
only for an SU(2) gauge theory --- the full SU(3) case will be studied later.

The scale $g^2\mu$ and the linear size of the nucleus $L$ are the only
physically relevant dimensional parameters of the classical EFT. Any
dimensional quantity $P$ well defined within the EFT can then be
written as $(g^2\mu)^d\,f_P (g^2\mu L)$, where $d$ is the dimension of
$P$. All the non--trivial physical information is therefore contained
in the dimensionless function $f_P (g^2\mu L)$.  On the lattice, $P$
will generally depend on the lattice spacing $a$; this dependence can
be removed by taking the continuum limit $a\rightarrow 0$.  Assuming
$g=2$, the physically relevant values of $g^2\mu$ for RHIC and LHC
energies are $\sim 2$ GeV and $\sim 4$ GeV respectively. Also,
assuming central Au--Au collisions, we obtain $L=11.6$ fm as the
physical linear dimension of our square lattice.  Thus, $g^2\mu L
\approx 120$ ($240$) for RHIC (LHC). In
Ref.~\cite{AlexRaj3}, we computed the initial energy density, per unit
area per unit rapidity, to be $dE/L^2/d\eta=f_E\,(g^2\mu)^3$.  We
computed $f_E$ as a function of $g^2\mu L$, and extrapolated our
results for $f_E$ to the continuum limit. We could however just as
well expressed our result as $dE/\pi R^2/d\eta = c_E(Q_s R)\,
(Q_s)^3$, with $c_E\sim 4.3$--$4.9$ in the region of
interest.

We will now report on our results for the initial multiplicity of
gluons produced at central rapidities in very high energy nuclear
collisions. This quantity, while not directly observable, is related
to the number of hadrons produced at central
rapidities~\cite{Larry1}. The initial multiplicity and momentum
distribution of gluons also determine the equilibration time, the
temperature and the chemical potential of the
QGP~\cite{Mueller,JeffRaj}.  The various signatures of QGP formation
are highly sensitive to these quantitites~\cite{QM99}.

We must first clarify what we mean by the number of quanta in the interacting,
non-Abelian gauge theory at hand.
To motivate the discussion, let us first consider
a free field theory whose Hamiltonian in momentum space has the form
\be
H_f = {1\over 2}\sum_k\,\left(|\pi(k)|^2 + \omega^2 (k)\,|\phi(k)|^2\right)\, ,
\ee
where $\phi(k)$ is the $k$th momentum component of the field, $\pi(k)$
is its conjugate momentum, and $\omega(k)$ is the corresponding
eigenfrequency.  The average particle number of the $k$-th mode is
then
\begin{equation}
N(k)=\omega(k)\langle|\phi(k)|^2\rangle
=\sqrt{\langle|\phi(k)|^2|\pi(k)|^2\rangle},\label{nfree} \, .
\end{equation}
In our case, the average $\langle\rangle$ is over the initial
conditions.

Clearly, any extension of this notion to interacting theories should
reduce to the standard free-field definition of the particle number in
the weak coupling limit. However, this requirement alone does not
define the particle number uniquely outside a free theory. We
therefore use two different generalizations of the particle number to
an interacting theory. Each has the correct free-field limit.  Even
though the fields in question are strongly interacting at early times, they are
only weakly coupled at late times, and it is only at this stage that
it becomes reasonable to define particle number.  We verify that the
two definitions agree in this weak-coupling regime.

Our first definition of the multiplicity is straightforward. We impose
the Coulomb gauge condition in the transverse plane,
${\vec\nabla}_\perp\cdot{\vec A}_\perp=0$, and substitute the momentum
components of the resulting field configuration into
Eq.~(\ref{nfree}). One option now is to assume $\omega(k_\perp)$ to be the
standard massless (lattice) dispersion relation and use the middle
expression of Eq.~(\ref{nfree}) to compute $N(k_\perp)$.  Alternatively, we
can determine $N(k_\perp)$ from the rightmost expression of
Eq.~(\ref{nfree}); the middle expression of Eq.~(\ref{nfree}) can then
be used to obtain $\omega(k_\perp)$. The second option is preferable; it
does not require us to assume that the dispersion relation is linear.

Our second definition is based on the behavior of a free-field theory under
cooling. Consider a simple relaxation equation for a field in real space,
\begin{equation}
\partial_t\phi(x)=-\partial H/\partial\phi(x),\label{relax}
\end{equation}
where $t$ is the cooling time ({\em not to be confused with real or
proper time}) and $H$ is the Hamiltonian. For a free field theory ($H=H_f$)
the relaxation equation has exactly the same form in the momentum
space with the solution $\phi(k,t)=\phi(k,0)\,{\rm
exp}(-\omega^2(k)t)$. The potential energy of the relaxed free field
is $V(t)=(1/2)\sum_k\omega^2(k)|\phi(k,t)|^2$.  It is then easy to
derive the following integral expression for the total particle number
of a free-field system:
\begin{equation}
N=\sqrt{8\over\pi}\int_0^\infty{{{\rm d}t}\over\sqrt{t}}\,V(t).\label{ncool}
\end{equation}
Eq.~(\ref{relax}) can be solved numerically for interacting fields.
Subsequently, $V(t)$ can be determined, and $N$ can be computed by
numerical integration. Note that in a gauge theory the relaxation
equations are gauge-covariant, and the relaxed potential $V(t)$ is
gauge-invariant, entailing gauge invariance of this definition of the
particle number.  This is an attractive feature of the cooling 
method. On the other hand, unlike the Coulomb gauge computation
discussed earlier, this cooling technique presently only permits
determination of the total particle number. It cannot be used to find
the number distribution $N(k_\perp)$.

Both our definitions cease to make sense if the system is far from
linearity.  In particular, if the theory has metastable states, and
the system relaxes to one of these states, the right hand side of
(\ref{ncool}) will diverge.  The expression in Eq.~(\ref{nfree}) can then
still be formally determined in the Coulomb gauge but its
interpretation as a particle number is problematic.  This situation
deserves special attention and will be discussed in detail
elsewhere. We did not observe any effects of metastabilty within the
range of parameters of the current numerical study. In particular, we
verified the convergence of Eq.~(\ref{ncool}) with respect to the
upper limit of integration.

We now present our results using both the techniques discussed. We
begin with the number distribution, which can only be computed in
Coulomb gauge.  We have verified, for $g^2\mu L=35.35$, that in the
range of values of $g^2\mu a$ considered here the system is close to
the continuum limit. This is consistent with our earlier analysis of
the lattice spacing dependence of a more ultraviolet-sensitive
quantity, the energy density \cite{AlexRaj3}.  In Fig.~1a, we plot the
gluon number distribution, $n(k_\perp)\equiv
dN/L^2/dk_\perp^2=N(k_\perp)/(2\pi)^2$ versus $k_\perp$ for fixed
$g^2\mu L=35.5$, but for different values of the lattice spacing
$g^2\mu a$.  For large $k_\perp$, the finest lattice ($g^2\mu a =
0.138$) agrees well with the lattice perturbation theory (LPTh)
analogue of Eq.~(\ref{LPT}). At smaller $k_\perp$, the distribution is
softer, and converges to a constant value.  In Fig.~1b, we plot the
gluon distribution in the infrared, at fixed $g^2\mu a$, for different
$g^2\mu L$ ($148.5$ and $297$). We notice that these distributions are
nearly universal and independent of $g^2\mu L$!  Also, the convergence
of the distribution to a constant value is more clearly visible in
Fig.~1b.

When $k_\perp\leq g^2\mu$, non--perturbative effects qualitatively alter
the perturbative number distribution, rendering it finite in the
infrared. Unfortunately, since these effects are large, an analytical
understanding of the behavior at low $k_\perp$ is lacking.  Our results,
despite being universal, are not simply fit by any of the physically
motivated parametrizations we have considered.

\begin{figure}[hbt]
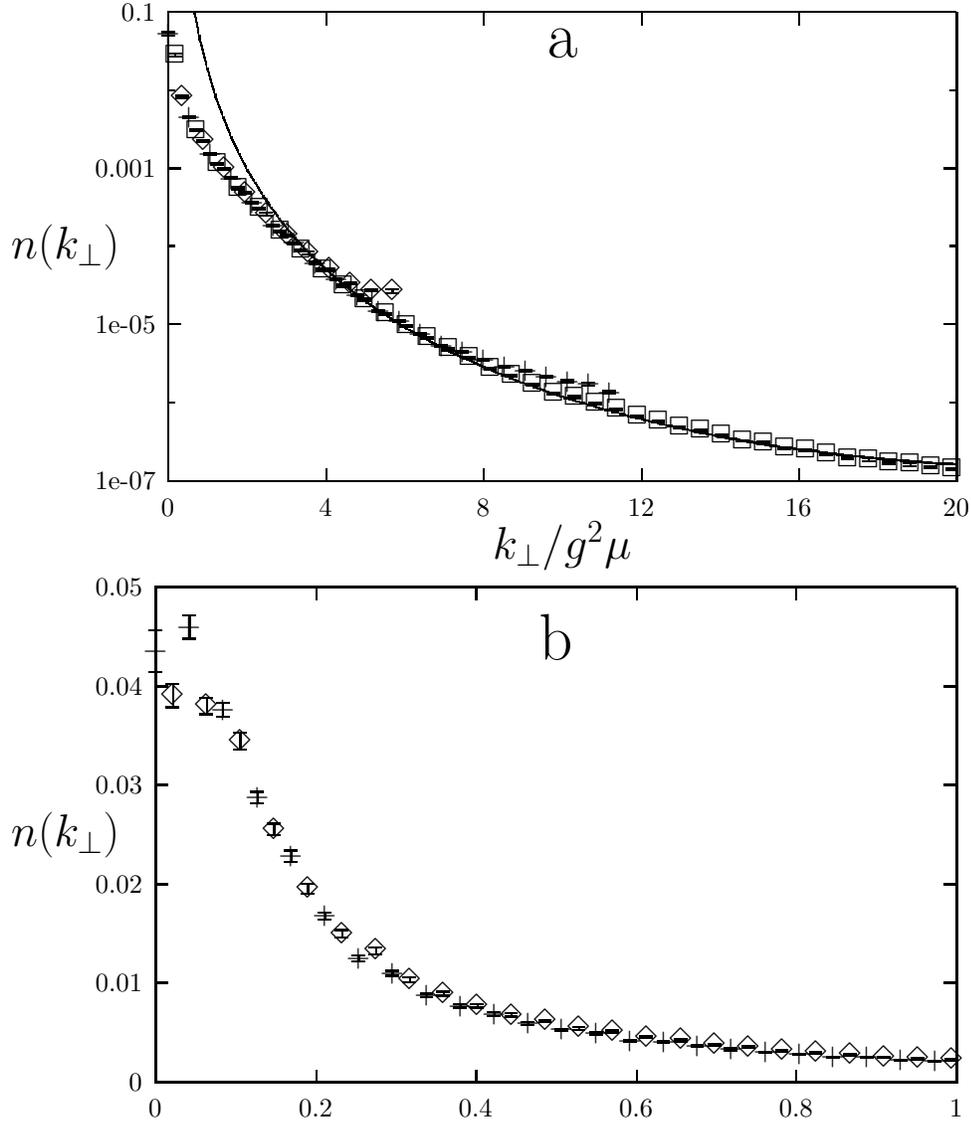

\input nvska
\input nvsk1
\caption{a: $n(k_\perp)\equiv dN/L^2/d^2 k_\perp$ as a function of the
gluon momentum $k$ for $g^2\mu L=35.35$ and the values 0.138
(squares), 0.276 (plusses), and 0.552 (diamonds) of $g^2\mu a$. The
gluon momentum $k$ is in units of $g^2\mu$. The solid line is a fit of
the lattice analog of the perturbative expression Eq.~(\ref{LPT}) to
the high-momentum part of the $g^2\mu a=0.138$ data.  b: $n(k_\perp)$
at soft momenta at $g^2\mu a=0.29$ for the values 148.5 (plusses) and
297 (diamonds) of $g^2\mu L$.} 
\label{nvsk}\end{figure}

From our previous discussion, the formula
\be
{1\over L^2}\,{dN\over d\eta} = {1\over g^2}\,
f_N(g^2\mu L)\, (g^2\mu)^2 \, ,
\label{BJN}
\ee
relates the number of produced gluons per unit area per unit rapidity
at zero rapidity to $g^2\mu$. We have computed $f_N$ on the lattice using
the two techniques discussed earlier. Our result for $f_N$ as a
function of $g^2\mu L$, for the smallest values of $g^2\mu a$
feasible, are plotted in Fig.~2. We see that the agreement between the
cooling and Coulomb techniques at larger values of $g^2\mu L$ is
excellent.  It is not as good at the smaller values -- in general, the
cooling number is more reliable~\cite{comment3}. We also note that
Fig.~2 demonstrates that the distributions in Fig.~1b are not quite
universal --- otherwise, $f_N$ would be a constant. We see instead that
it has a weak logarithmic rise with $g^2\mu L$ for larger $g^2\mu
L$'s.  Table~1 lists $f_N$ for various $g^2\mu L$. The third row
is the Coulomb gauge number after cooling --- see Ref.~\cite{comment3}.

\begin{figure}[hbt]
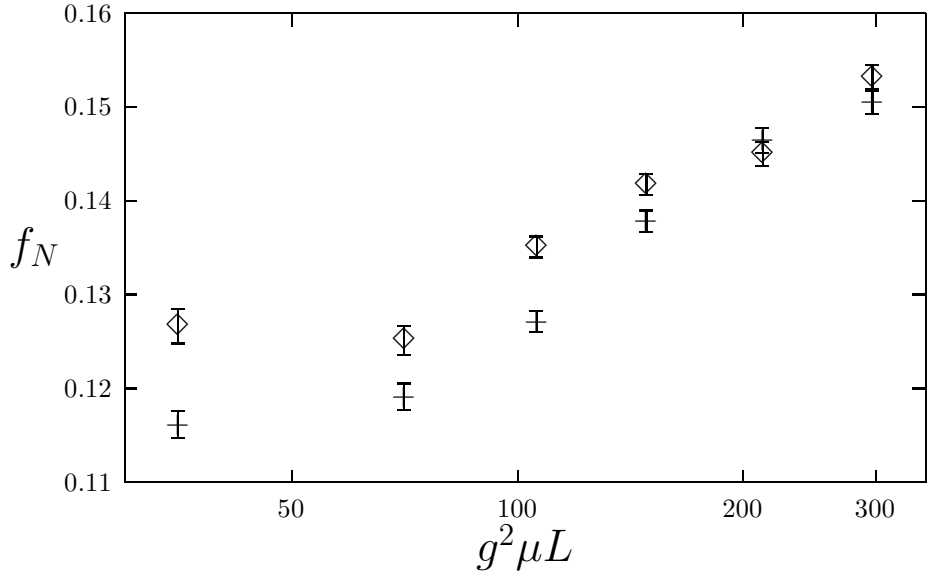

\input nvsmul
\caption{The function $f_N$, defined in Eq.~(\ref{BJN}) as a function of
$g^2\mu L$, obtained by the relaxation method (plusses) and by the Coulomb
gauge fixing (diamonds). The values of $g^2\mu a$ are 0.276 for
$g^2\mu L=35.35$ and $g^2\mu L=70.8$; 0.29 for $g^2\mu L=148.5$ and 
$g^2\mu L=297$; and 0.414 for $g^2\mu L=212$.} 
\label{nvsmul}\end{figure}
\begin{table}[h]
\begin{small}
\centerline{\begin{tabular}{|lrrrrrr|} \hline
$g^2\mu L$ & 35.36 & 70.71 & 106.1 & 148.5 & 212.1 & 297.0 \\
$g^2\mu a$  & .276 &  .276 & 0.207 & .29       & .41       & .29       \\
$f_N$ (cooling)  & $.116\pm .001$ & $.119\pm .001$ & $.127\pm .001$ & $.138\pm 
.001$ & 
$.146\pm .001$ & $.151\pm .001$   \\
$f_N$ (Coulomb) & $.127\pm .002$  & $.125\pm .002$ & $.135\pm 0.001$ & $142\pm 
.001$       & $.145\pm .001$  & $.153\pm .001$ \\
$f_N$ (res.)$\times 10^3$ & $14\pm 2$  & $7.8\pm 0.2$ & $8.9\pm 0.2$ & 
$5.6\pm 0.1$ & $7.12\pm 
0.08$  & $4.83\pm .04$ \\
\hline
\end{tabular}}
\caption{Values of $f_N$ vs $g^2\mu L$, for fixed $g^2\mu a$, 
plotted in Fig.~\ref{nvsmul}. $f_N$ (res.) is 
defined in Ref.~\cite{comment3}.}
\label{nvsmultab}
\end{small}
\end{table}
In Fig.~3, we plot the dispersion relation $\omega(k_\perp)$ vs
$k_\perp$ using the relation Eq.~(\ref{nfree}).  All the dispersion
curves rapidly approach the $\omega(k_\perp)=k_\perp$ asymptote
characteristic of on-shell partons, while exibiting a mass gap at zero
momentum.  We reserve for a later work a detailed study of this mass
gap and its role in rendering the number distributions infrared
finite.

\begin{figure}[hbt]
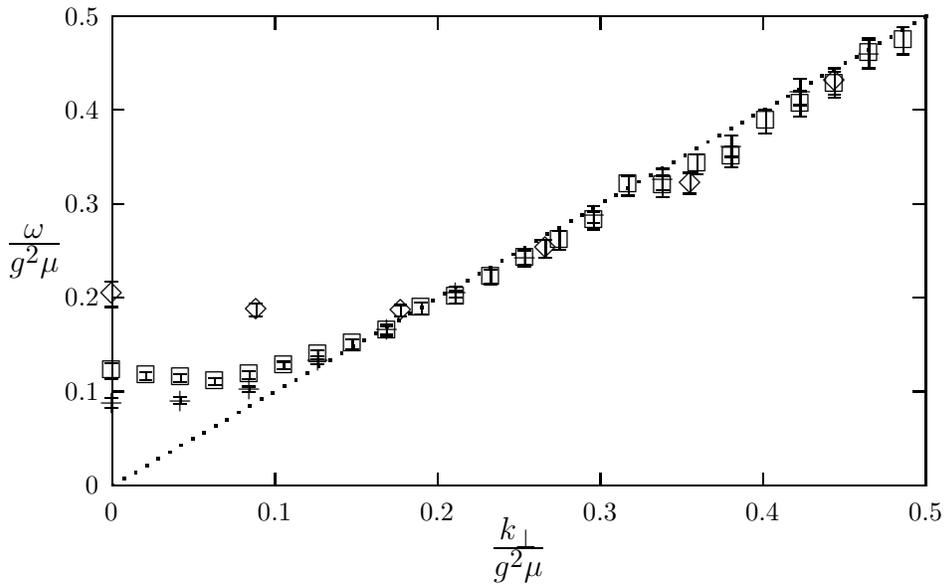

\input dr
\caption{Gluon dispersion relation $\omega(k_\perp)$ obtained from
Eq.~(\ref{nfree}), for the values 70.8 (diamonds), 148.5 (plusses),
and 297 (squares), with the values of $g^2\mu a$ as in Figure
\ref{nvsmul}.}
\label{dr}\end{figure}

We can compare our results for the number distribution to the one
predicted by A. H. Mueller~\cite{Mueller}. In terms of $Q_s$ and $R$,
we can re--write Eq.~\ref{BJN} as 
\be
\frac{1}{\pi R^2}\,\frac{dN}{d\eta}
= c_N\, \frac{N_c^2-1}{N_c}\,\frac{1}{4\pi^2\alpha_S}\,Q_s^2 \, .\nonumber
\ee
Mueller estimates the non--perturbative coefficient $c_N$ to be of
order unity. If we take $f_N=0.14\pm 0.01$, as is the case for much of
the range studied, we find $c_N=1.29\pm 0.09$, a number of order unity
as predicted by Mueller.  Despite this close agreement, the transverse
momentum distributions, shown in Fig.~1a and 1b, and discussed
earlier, look quite different from Mueller's guess of
$\theta(Q_s^2-k_\perp^2)$. The $\theta$--function distribution was only a
rough guess to represent a qualitative change of the distributions at
$k_\perp\sim Q_s$.

A large number of models of particle production in nuclear collisions
at RHIC and LHC energies can be found in the literature.  A nice
recent summary of their various predictions and relevant references
can be found in the compilation of
Ref.~\cite{Pajares}. Naively extrapolating our results to SU(3), we
find for $Au$-$Au$ central collisions at RHIC energies, $dN/d\eta
\sim 950$ for $f_N=0.132\pm 0.006$ ($g^2\mu L\approx 120$ --- 
we take the mean of
the $106$ and $148$ cooling point).  Similarly, for LHC energies $f_N=
0.148\pm 0.002$ ($g^2\mu L\sim 255$ --- the mean of the $212$ and
$297$ cooling point), one finds $dN/d\eta \sim 4300$.  In particular,
comparing our predictions with those of pQCD based models~\cite{pQCD},
we find our numbers to be in rough agreement. However, if we include a
$K$ factor like many of these models do, our numbers will be roughly a
factor of $2$ larger.

There is considerable uncertainity in the value of $Q_s$ because the
gluon densities at the relevant $x$ and $Q^2$ are ill--known.  Since
the multiplicity depends quadratically on $Q_s$, a prediction of the
same is perforce unreliable. Distinguishing between different models
will therefore require, at the very least, testing their predictions
for the scaling of multiplicities with $A$ and with $\sqrt{s}$. In our
case, $Q_s\sim A^{1/6}$, hence from Eq.~\ref{BJN} the number per
unit rapidity will, up to logarithms of $A$, be proportional to $A$.
In the EFT, one naively expects the energy dependence to be a power
law, the power being determined by the rise in the gluon density at
small $x$. Quantitative estimates of the dependence of $Q_s$ with
energy in the saturation region are being developed.  Predictions from
other models vary significantly, ranging from a power law
dependence~\cite{yurial} of $Q_s$ to $Q_s\propto
\exp(\sqrt{\ln(s/s_0)})$, where $s_0$ is a constant.  In the latter
case~\cite{LR}, it is claimed that a good fit to the multiplicity from
existing high energy hadron scattering data is obtained. Data from
RHIC will help constrain the energy dependence of $Q_s$.

The reader should also note that our relations are derived only for
the initial parton multiplicity distributions at central rapidities.
These provide the initial conditions for the subsequent evolution of
the system, which can be investigated in a transport
approach~\cite{Mueller,JeffRaj}. The rate of chemical equilibration,
and uncertainities due to hadronization also have to be taken into
account in predictions of observables such as charged hadron
multiplicities. Conversely, these observables may help constrain the 
saturation scale $Q_s$, and inform us about the very earliest 
stages of nuclear collisons.

In summary, we have derived a non--perturbative relation between the
multiplicity of produced partons and the saturation scale of parton
distributions in high energy nuclear collisions.  We have computed
number distributions which have the predicted perturbative 
behavior in the ultraviolet, and are finite in the infrared. At present, in
our approach, we are only able to make qualitative ``ball-park''
predictions. However, we have developed a framework in which these can
be quantified and extended in a consistent manner to study a large
number of final state observables in heavy ion collisions.

\section*{Acknowledgments}

We would like to thank Larry McLerran and Al Mueller for very useful
discussions.  R. V.'s research was supported by DOE Contract
No. DE-AC02-98CH10886.  The authors acknowledge support from the
Portuguese FCT, under grants CERN/P/FIS/1203/98 and
CERN/P/FIS/15196/1999.

\end{document}

%% file: nvska.tex
\setlength{\unitlength}{0.240900pt}
\ifx\plotpoint\undefined\newsavebox{\plotpoint}\fi
\sbox{\plotpoint}{\rule[-0.200pt]{0.400pt}{0.400pt}}%
\begin{picture}(1500,900)(0,0)
\font\gnuplot=cmr10 at 10pt
\gnuplot
\sbox{\plotpoint}{\rule[-0.200pt]{0.400pt}{0.400pt}}%
\put(201.0,123.0){\rule[-0.200pt]{4.818pt}{0.400pt}}
\put(181,123){\makebox(0,0)[r]{1e-07}}
\put(1419.0,123.0){\rule[-0.200pt]{4.818pt}{0.400pt}}
\put(201.0,246.0){\rule[-0.200pt]{2.409pt}{0.400pt}}
\put(1429.0,246.0){\rule[-0.200pt]{2.409pt}{0.400pt}}
\put(201.0,369.0){\rule[-0.200pt]{4.818pt}{0.400pt}}
\put(181,369){\makebox(0,0)[r]{1e-05}}
\put(1419.0,369.0){\rule[-0.200pt]{4.818pt}{0.400pt}}
\put(201.0,492.0){\rule[-0.200pt]{2.409pt}{0.400pt}}
\put(1429.0,492.0){\rule[-0.200pt]{2.409pt}{0.400pt}}
\put(201.0,614.0){\rule[-0.200pt]{4.818pt}{0.400pt}}
\put(181,614){\makebox(0,0)[r]{0.001}}
\put(1419.0,614.0){\rule[-0.200pt]{4.818pt}{0.400pt}}
\put(201.0,737.0){\rule[-0.200pt]{2.409pt}{0.400pt}}
\put(1429.0,737.0){\rule[-0.200pt]{2.409pt}{0.400pt}}
\put(201.0,860.0){\rule[-0.200pt]{4.818pt}{0.400pt}}
\put(181,860){\makebox(0,0)[r]{0.1}}
\put(1419.0,860.0){\rule[-0.200pt]{4.818pt}{0.400pt}}
\put(201.0,123.0){\rule[-0.200pt]{0.400pt}{4.818pt}}
\put(201,82){\makebox(0,0){0}}
\put(201.0,840.0){\rule[-0.200pt]{0.400pt}{4.818pt}}
\put(449.0,123.0){\rule[-0.200pt]{0.400pt}{4.818pt}}
\put(449,82){\makebox(0,0){4}}
\put(449.0,840.0){\rule[-0.200pt]{0.400pt}{4.818pt}}
\put(696.0,123.0){\rule[-0.200pt]{0.400pt}{4.818pt}}
\put(696,82){\makebox(0,0){8}}
\put(696.0,840.0){\rule[-0.200pt]{0.400pt}{4.818pt}}
\put(944.0,123.0){\rule[-0.200pt]{0.400pt}{4.818pt}}
\put(944,82){\makebox(0,0){12}}
\put(944.0,840.0){\rule[-0.200pt]{0.400pt}{4.818pt}}
\put(1191.0,123.0){\rule[-0.200pt]{0.400pt}{4.818pt}}
\put(1191,82){\makebox(0,0){16}}
\put(1191.0,840.0){\rule[-0.200pt]{0.400pt}{4.818pt}}
\put(1439.0,123.0){\rule[-0.200pt]{0.400pt}{4.818pt}}
\put(1439,82){\makebox(0,0){20}}
\put(1439.0,840.0){\rule[-0.200pt]{0.400pt}{4.818pt}}
\put(201.0,123.0){\rule[-0.200pt]{298.234pt}{0.400pt}}
\put(1439.0,123.0){\rule[-0.200pt]{0.400pt}{177.543pt}}
\put(201.0,860.0){\rule[-0.200pt]{298.234pt}{0.400pt}}
\put(40,491){\makebox(0,0){\Large $n(k_\perp)$}}
\put(820,21){\makebox(0,0){\Large ${k_\perp/{g^2\mu}}$}}
\put(820,811){\makebox(0,0){\Huge  a}}
\put(201.0,123.0){\rule[-0.200pt]{0.400pt}{177.543pt}}
\put(223.0,725.0){\rule[-0.200pt]{0.400pt}{0.723pt}}
\put(213.0,725.0){\rule[-0.200pt]{4.818pt}{0.400pt}}
\put(213.0,728.0){\rule[-0.200pt]{4.818pt}{0.400pt}}
\put(256.0,655.0){\rule[-0.200pt]{0.400pt}{0.723pt}}
\put(246.0,655.0){\rule[-0.200pt]{4.818pt}{0.400pt}}
\put(246.0,658.0){\rule[-0.200pt]{4.818pt}{0.400pt}}
\put(289.0,612.0){\rule[-0.200pt]{0.400pt}{0.723pt}}
\put(279.0,612.0){\rule[-0.200pt]{4.818pt}{0.400pt}}
\put(279.0,615.0){\rule[-0.200pt]{4.818pt}{0.400pt}}
\put(322.0,574.0){\rule[-0.200pt]{0.400pt}{0.723pt}}
\put(312.0,574.0){\rule[-0.200pt]{4.818pt}{0.400pt}}
\put(312.0,577.0){\rule[-0.200pt]{4.818pt}{0.400pt}}
\put(355.0,540.0){\rule[-0.200pt]{0.400pt}{0.964pt}}
\put(345.0,540.0){\rule[-0.200pt]{4.818pt}{0.400pt}}
\put(345.0,544.0){\rule[-0.200pt]{4.818pt}{0.400pt}}
\put(388.0,507.0){\rule[-0.200pt]{0.400pt}{0.723pt}}
\put(378.0,507.0){\rule[-0.200pt]{4.818pt}{0.400pt}}
\put(378.0,510.0){\rule[-0.200pt]{4.818pt}{0.400pt}}
\put(421.0,479.0){\rule[-0.200pt]{0.400pt}{0.964pt}}
\put(411.0,479.0){\rule[-0.200pt]{4.818pt}{0.400pt}}
\put(411.0,483.0){\rule[-0.200pt]{4.818pt}{0.400pt}}
\put(454.0,453.0){\rule[-0.200pt]{0.400pt}{0.964pt}}
\put(444.0,453.0){\rule[-0.200pt]{4.818pt}{0.400pt}}
\put(444.0,457.0){\rule[-0.200pt]{4.818pt}{0.400pt}}
\put(487.0,432.0){\rule[-0.200pt]{0.400pt}{0.723pt}}
\put(477.0,432.0){\rule[-0.200pt]{4.818pt}{0.400pt}}
\put(477.0,435.0){\rule[-0.200pt]{4.818pt}{0.400pt}}
\put(520.0,421.0){\rule[-0.200pt]{0.400pt}{0.723pt}}
\put(510.0,421.0){\rule[-0.200pt]{4.818pt}{0.400pt}}
\put(510.0,424.0){\rule[-0.200pt]{4.818pt}{0.400pt}}
\put(553.0,418.0){\rule[-0.200pt]{0.400pt}{1.445pt}}
\put(543.0,418.0){\rule[-0.200pt]{4.818pt}{0.400pt}}
\put(223,726){\raisebox{-.8pt}{\makebox(0,0){$\Diamond$}}}
\put(256,657){\raisebox{-.8pt}{\makebox(0,0){$\Diamond$}}}
\put(289,614){\raisebox{-.8pt}{\makebox(0,0){$\Diamond$}}}
\put(322,575){\raisebox{-.8pt}{\makebox(0,0){$\Diamond$}}}
\put(355,542){\raisebox{-.8pt}{\makebox(0,0){$\Diamond$}}}
\put(388,509){\raisebox{-.8pt}{\makebox(0,0){$\Diamond$}}}
\put(421,481){\raisebox{-.8pt}{\makebox(0,0){$\Diamond$}}}
\put(454,455){\raisebox{-.8pt}{\makebox(0,0){$\Diamond$}}}
\put(487,433){\raisebox{-.8pt}{\makebox(0,0){$\Diamond$}}}
\put(520,422){\raisebox{-.8pt}{\makebox(0,0){$\Diamond$}}}
\put(553,421){\raisebox{-.8pt}{\makebox(0,0){$\Diamond$}}}
\put(543.0,424.0){\rule[-0.200pt]{4.818pt}{0.400pt}}
\put(201.0,823.0){\rule[-0.200pt]{0.400pt}{1.204pt}}
\put(191.0,823.0){\rule[-0.200pt]{4.818pt}{0.400pt}}
\put(191.0,828.0){\rule[-0.200pt]{4.818pt}{0.400pt}}
\put(234.0,693.0){\rule[-0.200pt]{0.400pt}{0.723pt}}
\put(224.0,693.0){\rule[-0.200pt]{4.818pt}{0.400pt}}
\put(224.0,696.0){\rule[-0.200pt]{4.818pt}{0.400pt}}
\put(267.0,635.0){\rule[-0.200pt]{0.400pt}{0.723pt}}
\put(257.0,635.0){\rule[-0.200pt]{4.818pt}{0.400pt}}
\put(257.0,638.0){\rule[-0.200pt]{4.818pt}{0.400pt}}
\put(300.0,597.0){\rule[-0.200pt]{0.400pt}{0.723pt}}
\put(290.0,597.0){\rule[-0.200pt]{4.818pt}{0.400pt}}
\put(290.0,600.0){\rule[-0.200pt]{4.818pt}{0.400pt}}
\put(333.0,559.0){\rule[-0.200pt]{0.400pt}{0.723pt}}
\put(323.0,559.0){\rule[-0.200pt]{4.818pt}{0.400pt}}
\put(323.0,562.0){\rule[-0.200pt]{4.818pt}{0.400pt}}
\put(366.0,522.0){\rule[-0.200pt]{0.400pt}{0.723pt}}
\put(356.0,522.0){\rule[-0.200pt]{4.818pt}{0.400pt}}
\put(356.0,525.0){\rule[-0.200pt]{4.818pt}{0.400pt}}
\put(399.0,495.0){\rule[-0.200pt]{0.400pt}{0.723pt}}
\put(389.0,495.0){\rule[-0.200pt]{4.818pt}{0.400pt}}
\put(389.0,498.0){\rule[-0.200pt]{4.818pt}{0.400pt}}
\put(432.0,464.0){\rule[-0.200pt]{0.400pt}{0.723pt}}
\put(422.0,464.0){\rule[-0.200pt]{4.818pt}{0.400pt}}
\put(422.0,467.0){\rule[-0.200pt]{4.818pt}{0.400pt}}
\put(465.0,438.0){\rule[-0.200pt]{0.400pt}{0.723pt}}
\put(455.0,438.0){\rule[-0.200pt]{4.818pt}{0.400pt}}
\put(455.0,441.0){\rule[-0.200pt]{4.818pt}{0.400pt}}
\put(498.0,413.0){\rule[-0.200pt]{0.400pt}{0.723pt}}
\put(488.0,413.0){\rule[-0.200pt]{4.818pt}{0.400pt}}
\put(488.0,416.0){\rule[-0.200pt]{4.818pt}{0.400pt}}
\put(531.0,389.0){\rule[-0.200pt]{0.400pt}{0.482pt}}
\put(521.0,389.0){\rule[-0.200pt]{4.818pt}{0.400pt}}
\put(521.0,391.0){\rule[-0.200pt]{4.818pt}{0.400pt}}
\put(564.0,372.0){\rule[-0.200pt]{0.400pt}{0.964pt}}
\put(554.0,372.0){\rule[-0.200pt]{4.818pt}{0.400pt}}
\put(554.0,376.0){\rule[-0.200pt]{4.818pt}{0.400pt}}
\put(597.0,353.0){\rule[-0.200pt]{0.400pt}{0.482pt}}
\put(587.0,353.0){\rule[-0.200pt]{4.818pt}{0.400pt}}
\put(587.0,355.0){\rule[-0.200pt]{4.818pt}{0.400pt}}
\put(630.0,334.0){\rule[-0.200pt]{0.400pt}{0.723pt}}
\put(620.0,334.0){\rule[-0.200pt]{4.818pt}{0.400pt}}
\put(620.0,337.0){\rule[-0.200pt]{4.818pt}{0.400pt}}
\put(663.0,324.0){\rule[-0.200pt]{0.400pt}{0.723pt}}
\put(653.0,324.0){\rule[-0.200pt]{4.818pt}{0.400pt}}
\put(653.0,327.0){\rule[-0.200pt]{4.818pt}{0.400pt}}
\put(696.0,311.0){\rule[-0.200pt]{0.400pt}{0.723pt}}
\put(686.0,311.0){\rule[-0.200pt]{4.818pt}{0.400pt}}
\put(686.0,314.0){\rule[-0.200pt]{4.818pt}{0.400pt}}
\put(729.0,301.0){\rule[-0.200pt]{0.400pt}{0.723pt}}
\put(719.0,301.0){\rule[-0.200pt]{4.818pt}{0.400pt}}
\put(719.0,304.0){\rule[-0.200pt]{4.818pt}{0.400pt}}
\put(762.0,294.0){\rule[-0.200pt]{0.400pt}{0.723pt}}
\put(752.0,294.0){\rule[-0.200pt]{4.818pt}{0.400pt}}
\put(752.0,297.0){\rule[-0.200pt]{4.818pt}{0.400pt}}
\put(795.0,285.0){\rule[-0.200pt]{0.400pt}{0.723pt}}
\put(785.0,285.0){\rule[-0.200pt]{4.818pt}{0.400pt}}
\put(785.0,288.0){\rule[-0.200pt]{4.818pt}{0.400pt}}
\put(828.0,277.0){\rule[-0.200pt]{0.400pt}{0.964pt}}
\put(818.0,277.0){\rule[-0.200pt]{4.818pt}{0.400pt}}
\put(818.0,281.0){\rule[-0.200pt]{4.818pt}{0.400pt}}
\put(861.0,273.0){\rule[-0.200pt]{0.400pt}{0.964pt}}
\put(851.0,273.0){\rule[-0.200pt]{4.818pt}{0.400pt}}
\put(851.0,277.0){\rule[-0.200pt]{4.818pt}{0.400pt}}
\put(894.0,260.0){\rule[-0.200pt]{0.400pt}{0.964pt}}
\put(884.0,260.0){\rule[-0.200pt]{4.818pt}{0.400pt}}
\put(201,826){\makebox(0,0){$+$}}
\put(234,695){\makebox(0,0){$+$}}
\put(267,637){\makebox(0,0){$+$}}
\put(300,598){\makebox(0,0){$+$}}
\put(333,560){\makebox(0,0){$+$}}
\put(366,524){\makebox(0,0){$+$}}
\put(399,497){\makebox(0,0){$+$}}
\put(432,465){\makebox(0,0){$+$}}
\put(465,440){\makebox(0,0){$+$}}
\put(498,415){\makebox(0,0){$+$}}
\put(531,390){\makebox(0,0){$+$}}
\put(564,374){\makebox(0,0){$+$}}
\put(597,354){\makebox(0,0){$+$}}
\put(630,335){\makebox(0,0){$+$}}
\put(663,325){\makebox(0,0){$+$}}
\put(696,313){\makebox(0,0){$+$}}
\put(729,302){\makebox(0,0){$+$}}
\put(762,296){\makebox(0,0){$+$}}
\put(795,287){\makebox(0,0){$+$}}
\put(828,279){\makebox(0,0){$+$}}
\put(861,275){\makebox(0,0){$+$}}
\put(894,262){\makebox(0,0){$+$}}
\put(884.0,264.0){\rule[-0.200pt]{4.818pt}{0.400pt}}
\put(212.0,790.0){\rule[-0.200pt]{0.400pt}{0.964pt}}
\put(202.0,790.0){\rule[-0.200pt]{4.818pt}{0.400pt}}
\put(202.0,794.0){\rule[-0.200pt]{4.818pt}{0.400pt}}
\put(245.0,672.0){\rule[-0.200pt]{0.400pt}{0.723pt}}
\put(235.0,672.0){\rule[-0.200pt]{4.818pt}{0.400pt}}
\put(235.0,675.0){\rule[-0.200pt]{4.818pt}{0.400pt}}
\put(278.0,620.0){\rule[-0.200pt]{0.400pt}{0.723pt}}
\put(268.0,620.0){\rule[-0.200pt]{4.818pt}{0.400pt}}
\put(268.0,623.0){\rule[-0.200pt]{4.818pt}{0.400pt}}
\put(311.0,581.0){\rule[-0.200pt]{0.400pt}{0.723pt}}
\put(301.0,581.0){\rule[-0.200pt]{4.818pt}{0.400pt}}
\put(301.0,584.0){\rule[-0.200pt]{4.818pt}{0.400pt}}
\put(344.0,549.0){\rule[-0.200pt]{0.400pt}{0.723pt}}
\put(334.0,549.0){\rule[-0.200pt]{4.818pt}{0.400pt}}
\put(334.0,552.0){\rule[-0.200pt]{4.818pt}{0.400pt}}
\put(377.0,513.0){\rule[-0.200pt]{0.400pt}{0.723pt}}
\put(367.0,513.0){\rule[-0.200pt]{4.818pt}{0.400pt}}
\put(367.0,516.0){\rule[-0.200pt]{4.818pt}{0.400pt}}
\put(410.0,484.0){\rule[-0.200pt]{0.400pt}{0.723pt}}
\put(400.0,484.0){\rule[-0.200pt]{4.818pt}{0.400pt}}
\put(400.0,487.0){\rule[-0.200pt]{4.818pt}{0.400pt}}
\put(443.0,453.0){\rule[-0.200pt]{0.400pt}{0.723pt}}
\put(433.0,453.0){\rule[-0.200pt]{4.818pt}{0.400pt}}
\put(433.0,456.0){\rule[-0.200pt]{4.818pt}{0.400pt}}
\put(476.0,428.0){\rule[-0.200pt]{0.400pt}{0.723pt}}
\put(466.0,428.0){\rule[-0.200pt]{4.818pt}{0.400pt}}
\put(466.0,431.0){\rule[-0.200pt]{4.818pt}{0.400pt}}
\put(509.0,406.0){\rule[-0.200pt]{0.400pt}{0.723pt}}
\put(499.0,406.0){\rule[-0.200pt]{4.818pt}{0.400pt}}
\put(499.0,409.0){\rule[-0.200pt]{4.818pt}{0.400pt}}
\put(542.0,383.0){\rule[-0.200pt]{0.400pt}{0.723pt}}
\put(532.0,383.0){\rule[-0.200pt]{4.818pt}{0.400pt}}
\put(532.0,386.0){\rule[-0.200pt]{4.818pt}{0.400pt}}
\put(575.0,365.0){\rule[-0.200pt]{0.400pt}{0.723pt}}
\put(565.0,365.0){\rule[-0.200pt]{4.818pt}{0.400pt}}
\put(565.0,368.0){\rule[-0.200pt]{4.818pt}{0.400pt}}
\put(608.0,346.0){\rule[-0.200pt]{0.400pt}{0.723pt}}
\put(598.0,346.0){\rule[-0.200pt]{4.818pt}{0.400pt}}
\put(598.0,349.0){\rule[-0.200pt]{4.818pt}{0.400pt}}
\put(641.0,328.0){\rule[-0.200pt]{0.400pt}{0.723pt}}
\put(631.0,328.0){\rule[-0.200pt]{4.818pt}{0.400pt}}
\put(631.0,331.0){\rule[-0.200pt]{4.818pt}{0.400pt}}
\put(674.0,315.0){\rule[-0.200pt]{0.400pt}{0.482pt}}
\put(664.0,315.0){\rule[-0.200pt]{4.818pt}{0.400pt}}
\put(664.0,317.0){\rule[-0.200pt]{4.818pt}{0.400pt}}
\put(707.0,297.0){\rule[-0.200pt]{0.400pt}{0.723pt}}
\put(697.0,297.0){\rule[-0.200pt]{4.818pt}{0.400pt}}
\put(697.0,300.0){\rule[-0.200pt]{4.818pt}{0.400pt}}
\put(740.0,287.0){\rule[-0.200pt]{0.400pt}{0.723pt}}
\put(730.0,287.0){\rule[-0.200pt]{4.818pt}{0.400pt}}
\put(730.0,290.0){\rule[-0.200pt]{4.818pt}{0.400pt}}
\put(773.0,273.0){\rule[-0.200pt]{0.400pt}{0.723pt}}
\put(763.0,273.0){\rule[-0.200pt]{4.818pt}{0.400pt}}
\put(763.0,276.0){\rule[-0.200pt]{4.818pt}{0.400pt}}
\put(806.0,259.0){\rule[-0.200pt]{0.400pt}{0.723pt}}
\put(796.0,259.0){\rule[-0.200pt]{4.818pt}{0.400pt}}
\put(796.0,262.0){\rule[-0.200pt]{4.818pt}{0.400pt}}
\put(839.0,253.0){\rule[-0.200pt]{0.400pt}{0.723pt}}
\put(829.0,253.0){\rule[-0.200pt]{4.818pt}{0.400pt}}
\put(829.0,256.0){\rule[-0.200pt]{4.818pt}{0.400pt}}
\put(872.0,243.0){\rule[-0.200pt]{0.400pt}{0.723pt}}
\put(862.0,243.0){\rule[-0.200pt]{4.818pt}{0.400pt}}
\put(862.0,246.0){\rule[-0.200pt]{4.818pt}{0.400pt}}
\put(905.0,233.0){\rule[-0.200pt]{0.400pt}{0.723pt}}
\put(895.0,233.0){\rule[-0.200pt]{4.818pt}{0.400pt}}
\put(895.0,236.0){\rule[-0.200pt]{4.818pt}{0.400pt}}
\put(938.0,223.0){\rule[-0.200pt]{0.400pt}{0.723pt}}
\put(928.0,223.0){\rule[-0.200pt]{4.818pt}{0.400pt}}
\put(928.0,226.0){\rule[-0.200pt]{4.818pt}{0.400pt}}
\put(971.0,215.0){\rule[-0.200pt]{0.400pt}{0.482pt}}
\put(961.0,215.0){\rule[-0.200pt]{4.818pt}{0.400pt}}
\put(961.0,217.0){\rule[-0.200pt]{4.818pt}{0.400pt}}
\put(1004.0,205.0){\rule[-0.200pt]{0.400pt}{0.723pt}}
\put(994.0,205.0){\rule[-0.200pt]{4.818pt}{0.400pt}}
\put(994.0,208.0){\rule[-0.200pt]{4.818pt}{0.400pt}}
\put(1037.0,200.0){\rule[-0.200pt]{0.400pt}{0.723pt}}
\put(1027.0,200.0){\rule[-0.200pt]{4.818pt}{0.400pt}}
\put(1027.0,203.0){\rule[-0.200pt]{4.818pt}{0.400pt}}
\put(1070.0,193.0){\rule[-0.200pt]{0.400pt}{0.723pt}}
\put(1060.0,193.0){\rule[-0.200pt]{4.818pt}{0.400pt}}
\put(1060.0,196.0){\rule[-0.200pt]{4.818pt}{0.400pt}}
\put(1103.0,185.0){\rule[-0.200pt]{0.400pt}{0.482pt}}
\put(1093.0,185.0){\rule[-0.200pt]{4.818pt}{0.400pt}}
\put(1093.0,187.0){\rule[-0.200pt]{4.818pt}{0.400pt}}
\put(1136.0,181.0){\rule[-0.200pt]{0.400pt}{0.723pt}}
\put(1126.0,181.0){\rule[-0.200pt]{4.818pt}{0.400pt}}
\put(1126.0,184.0){\rule[-0.200pt]{4.818pt}{0.400pt}}
\put(1169.0,172.0){\rule[-0.200pt]{0.400pt}{0.723pt}}
\put(1159.0,172.0){\rule[-0.200pt]{4.818pt}{0.400pt}}
\put(1159.0,175.0){\rule[-0.200pt]{4.818pt}{0.400pt}}
\put(1202.0,169.0){\rule[-0.200pt]{0.400pt}{0.723pt}}
\put(1192.0,169.0){\rule[-0.200pt]{4.818pt}{0.400pt}}
\put(1192.0,172.0){\rule[-0.200pt]{4.818pt}{0.400pt}}
\put(1235.0,163.0){\rule[-0.200pt]{0.400pt}{0.723pt}}
\put(1225.0,163.0){\rule[-0.200pt]{4.818pt}{0.400pt}}
\put(1225.0,166.0){\rule[-0.200pt]{4.818pt}{0.400pt}}
\put(1268.0,157.0){\rule[-0.200pt]{0.400pt}{0.723pt}}
\put(1258.0,157.0){\rule[-0.200pt]{4.818pt}{0.400pt}}
\put(1258.0,160.0){\rule[-0.200pt]{4.818pt}{0.400pt}}
\put(1301.0,154.0){\rule[-0.200pt]{0.400pt}{0.964pt}}
\put(1291.0,154.0){\rule[-0.200pt]{4.818pt}{0.400pt}}
\put(1291.0,158.0){\rule[-0.200pt]{4.818pt}{0.400pt}}
\put(1333.0,150.0){\rule[-0.200pt]{0.400pt}{0.723pt}}
\put(1323.0,150.0){\rule[-0.200pt]{4.818pt}{0.400pt}}
\put(1323.0,153.0){\rule[-0.200pt]{4.818pt}{0.400pt}}
\put(1366.0,146.0){\rule[-0.200pt]{0.400pt}{1.204pt}}
\put(1356.0,146.0){\rule[-0.200pt]{4.818pt}{0.400pt}}
\put(1356.0,151.0){\rule[-0.200pt]{4.818pt}{0.400pt}}
\put(1399.0,143.0){\rule[-0.200pt]{0.400pt}{0.723pt}}
\put(1389.0,143.0){\rule[-0.200pt]{4.818pt}{0.400pt}}
\put(1389.0,146.0){\rule[-0.200pt]{4.818pt}{0.400pt}}
\put(1432.0,140.0){\rule[-0.200pt]{0.400pt}{0.723pt}}
\put(1422.0,140.0){\rule[-0.200pt]{4.818pt}{0.400pt}}
\put(212,792){\raisebox{-.8pt}{\makebox(0,0){$\Box$}}}
\put(245,673){\raisebox{-.8pt}{\makebox(0,0){$\Box$}}}
\put(278,621){\raisebox{-.8pt}{\makebox(0,0){$\Box$}}}
\put(311,583){\raisebox{-.8pt}{\makebox(0,0){$\Box$}}}
\put(344,551){\raisebox{-.8pt}{\makebox(0,0){$\Box$}}}
\put(377,515){\raisebox{-.8pt}{\makebox(0,0){$\Box$}}}
\put(410,485){\raisebox{-.8pt}{\makebox(0,0){$\Box$}}}
\put(443,454){\raisebox{-.8pt}{\makebox(0,0){$\Box$}}}
\put(476,429){\raisebox{-.8pt}{\makebox(0,0){$\Box$}}}
\put(509,408){\raisebox{-.8pt}{\makebox(0,0){$\Box$}}}
\put(542,385){\raisebox{-.8pt}{\makebox(0,0){$\Box$}}}
\put(575,366){\raisebox{-.8pt}{\makebox(0,0){$\Box$}}}
\put(608,348){\raisebox{-.8pt}{\makebox(0,0){$\Box$}}}
\put(641,330){\raisebox{-.8pt}{\makebox(0,0){$\Box$}}}
\put(674,316){\raisebox{-.8pt}{\makebox(0,0){$\Box$}}}
\put(707,299){\raisebox{-.8pt}{\makebox(0,0){$\Box$}}}
\put(740,289){\raisebox{-.8pt}{\makebox(0,0){$\Box$}}}
\put(773,275){\raisebox{-.8pt}{\makebox(0,0){$\Box$}}}
\put(806,261){\raisebox{-.8pt}{\makebox(0,0){$\Box$}}}
\put(839,254){\raisebox{-.8pt}{\makebox(0,0){$\Box$}}}
\put(872,244){\raisebox{-.8pt}{\makebox(0,0){$\Box$}}}
\put(905,235){\raisebox{-.8pt}{\makebox(0,0){$\Box$}}}
\put(938,224){\raisebox{-.8pt}{\makebox(0,0){$\Box$}}}
\put(971,216){\raisebox{-.8pt}{\makebox(0,0){$\Box$}}}
\put(1004,207){\raisebox{-.8pt}{\makebox(0,0){$\Box$}}}
\put(1037,202){\raisebox{-.8pt}{\makebox(0,0){$\Box$}}}
\put(1070,194){\raisebox{-.8pt}{\makebox(0,0){$\Box$}}}
\put(1103,186){\raisebox{-.8pt}{\makebox(0,0){$\Box$}}}
\put(1136,182){\raisebox{-.8pt}{\makebox(0,0){$\Box$}}}
\put(1169,174){\raisebox{-.8pt}{\makebox(0,0){$\Box$}}}
\put(1202,171){\raisebox{-.8pt}{\makebox(0,0){$\Box$}}}
\put(1235,165){\raisebox{-.8pt}{\makebox(0,0){$\Box$}}}
\put(1268,158){\raisebox{-.8pt}{\makebox(0,0){$\Box$}}}
\put(1301,156){\raisebox{-.8pt}{\makebox(0,0){$\Box$}}}
\put(1333,151){\raisebox{-.8pt}{\makebox(0,0){$\Box$}}}
\put(1366,149){\raisebox{-.8pt}{\makebox(0,0){$\Box$}}}
\put(1399,145){\raisebox{-.8pt}{\makebox(0,0){$\Box$}}}
\put(1432,141){\raisebox{-.8pt}{\makebox(0,0){$\Box$}}}
\put(1422.0,143.0){\rule[-0.200pt]{4.818pt}{0.400pt}}
\multiput(243.59,851.08)(0.488,-2.673){13}{\rule{0.117pt}{2.150pt}}
\multiput(242.17,855.54)(8.000,-36.538){2}{\rule{0.400pt}{1.075pt}}
\multiput(251.58,812.07)(0.493,-2.003){23}{\rule{0.119pt}{1.669pt}}
\multiput(250.17,815.54)(13.000,-47.535){2}{\rule{0.400pt}{0.835pt}}
\multiput(264.58,761.91)(0.492,-1.746){21}{\rule{0.119pt}{1.467pt}}
\multiput(263.17,764.96)(12.000,-37.956){2}{\rule{0.400pt}{0.733pt}}
\multiput(276.58,722.11)(0.493,-1.369){23}{\rule{0.119pt}{1.177pt}}
\multiput(275.17,724.56)(13.000,-32.557){2}{\rule{0.400pt}{0.588pt}}
\multiput(289.58,687.43)(0.492,-1.272){21}{\rule{0.119pt}{1.100pt}}
\multiput(288.17,689.72)(12.000,-27.717){2}{\rule{0.400pt}{0.550pt}}
\multiput(301.58,658.14)(0.493,-1.052){23}{\rule{0.119pt}{0.931pt}}
\multiput(300.17,660.07)(13.000,-25.068){2}{\rule{0.400pt}{0.465pt}}
\multiput(314.58,631.26)(0.492,-1.013){21}{\rule{0.119pt}{0.900pt}}
\multiput(313.17,633.13)(12.000,-22.132){2}{\rule{0.400pt}{0.450pt}}
\multiput(326.58,607.90)(0.493,-0.814){23}{\rule{0.119pt}{0.746pt}}
\multiput(325.17,609.45)(13.000,-19.451){2}{\rule{0.400pt}{0.373pt}}
\multiput(339.58,586.82)(0.492,-0.841){21}{\rule{0.119pt}{0.767pt}}
\multiput(338.17,588.41)(12.000,-18.409){2}{\rule{0.400pt}{0.383pt}}
\multiput(351.58,567.29)(0.493,-0.695){23}{\rule{0.119pt}{0.654pt}}
\multiput(350.17,568.64)(13.000,-16.643){2}{\rule{0.400pt}{0.327pt}}
\multiput(364.58,549.23)(0.492,-0.712){21}{\rule{0.119pt}{0.667pt}}
\multiput(363.17,550.62)(12.000,-15.616){2}{\rule{0.400pt}{0.333pt}}
\multiput(376.58,532.54)(0.493,-0.616){23}{\rule{0.119pt}{0.592pt}}
\multiput(375.17,533.77)(13.000,-14.771){2}{\rule{0.400pt}{0.296pt}}
\multiput(389.58,516.51)(0.492,-0.625){21}{\rule{0.119pt}{0.600pt}}
\multiput(388.17,517.75)(12.000,-13.755){2}{\rule{0.400pt}{0.300pt}}
\multiput(401.58,501.80)(0.493,-0.536){23}{\rule{0.119pt}{0.531pt}}
\multiput(400.17,502.90)(13.000,-12.898){2}{\rule{0.400pt}{0.265pt}}
\multiput(414.58,487.79)(0.492,-0.539){21}{\rule{0.119pt}{0.533pt}}
\multiput(413.17,488.89)(12.000,-11.893){2}{\rule{0.400pt}{0.267pt}}
\multiput(426.00,475.92)(0.539,-0.492){21}{\rule{0.533pt}{0.119pt}}
\multiput(426.00,476.17)(11.893,-12.000){2}{\rule{0.267pt}{0.400pt}}
\multiput(439.00,463.92)(0.543,-0.492){19}{\rule{0.536pt}{0.118pt}}
\multiput(439.00,464.17)(10.887,-11.000){2}{\rule{0.268pt}{0.400pt}}
\multiput(451.00,452.92)(0.539,-0.492){21}{\rule{0.533pt}{0.119pt}}
\multiput(451.00,453.17)(11.893,-12.000){2}{\rule{0.267pt}{0.400pt}}
\multiput(464.00,440.92)(0.600,-0.491){17}{\rule{0.580pt}{0.118pt}}
\multiput(464.00,441.17)(10.796,-10.000){2}{\rule{0.290pt}{0.400pt}}
\multiput(476.00,430.92)(0.652,-0.491){17}{\rule{0.620pt}{0.118pt}}
\multiput(476.00,431.17)(11.713,-10.000){2}{\rule{0.310pt}{0.400pt}}
\multiput(489.00,420.92)(0.600,-0.491){17}{\rule{0.580pt}{0.118pt}}
\multiput(489.00,421.17)(10.796,-10.000){2}{\rule{0.290pt}{0.400pt}}
\multiput(501.00,410.93)(0.728,-0.489){15}{\rule{0.678pt}{0.118pt}}
\multiput(501.00,411.17)(11.593,-9.000){2}{\rule{0.339pt}{0.400pt}}
\multiput(514.00,401.93)(0.669,-0.489){15}{\rule{0.633pt}{0.118pt}}
\multiput(514.00,402.17)(10.685,-9.000){2}{\rule{0.317pt}{0.400pt}}
\multiput(526.00,392.93)(0.824,-0.488){13}{\rule{0.750pt}{0.117pt}}
\multiput(526.00,393.17)(11.443,-8.000){2}{\rule{0.375pt}{0.400pt}}
\multiput(539.00,384.93)(0.758,-0.488){13}{\rule{0.700pt}{0.117pt}}
\multiput(539.00,385.17)(10.547,-8.000){2}{\rule{0.350pt}{0.400pt}}
\multiput(551.00,376.93)(0.824,-0.488){13}{\rule{0.750pt}{0.117pt}}
\multiput(551.00,377.17)(11.443,-8.000){2}{\rule{0.375pt}{0.400pt}}
\multiput(564.00,368.93)(0.758,-0.488){13}{\rule{0.700pt}{0.117pt}}
\multiput(564.00,369.17)(10.547,-8.000){2}{\rule{0.350pt}{0.400pt}}
\multiput(576.00,360.93)(0.950,-0.485){11}{\rule{0.843pt}{0.117pt}}
\multiput(576.00,361.17)(11.251,-7.000){2}{\rule{0.421pt}{0.400pt}}
\multiput(589.00,353.93)(0.874,-0.485){11}{\rule{0.786pt}{0.117pt}}
\multiput(589.00,354.17)(10.369,-7.000){2}{\rule{0.393pt}{0.400pt}}
\multiput(601.00,346.93)(0.950,-0.485){11}{\rule{0.843pt}{0.117pt}}
\multiput(601.00,347.17)(11.251,-7.000){2}{\rule{0.421pt}{0.400pt}}
\multiput(614.00,339.93)(1.033,-0.482){9}{\rule{0.900pt}{0.116pt}}
\multiput(614.00,340.17)(10.132,-6.000){2}{\rule{0.450pt}{0.400pt}}
\multiput(626.00,333.93)(0.950,-0.485){11}{\rule{0.843pt}{0.117pt}}
\multiput(626.00,334.17)(11.251,-7.000){2}{\rule{0.421pt}{0.400pt}}
\multiput(639.00,326.93)(1.033,-0.482){9}{\rule{0.900pt}{0.116pt}}
\multiput(639.00,327.17)(10.132,-6.000){2}{\rule{0.450pt}{0.400pt}}
\multiput(651.00,320.93)(1.123,-0.482){9}{\rule{0.967pt}{0.116pt}}
\multiput(651.00,321.17)(10.994,-6.000){2}{\rule{0.483pt}{0.400pt}}
\multiput(664.00,314.93)(1.267,-0.477){7}{\rule{1.060pt}{0.115pt}}
\multiput(664.00,315.17)(9.800,-5.000){2}{\rule{0.530pt}{0.400pt}}
\multiput(676.00,309.93)(1.123,-0.482){9}{\rule{0.967pt}{0.116pt}}
\multiput(676.00,310.17)(10.994,-6.000){2}{\rule{0.483pt}{0.400pt}}
\multiput(689.00,303.93)(1.267,-0.477){7}{\rule{1.060pt}{0.115pt}}
\multiput(689.00,304.17)(9.800,-5.000){2}{\rule{0.530pt}{0.400pt}}
\multiput(701.00,298.93)(1.123,-0.482){9}{\rule{0.967pt}{0.116pt}}
\multiput(701.00,299.17)(10.994,-6.000){2}{\rule{0.483pt}{0.400pt}}
\multiput(714.00,292.93)(1.267,-0.477){7}{\rule{1.060pt}{0.115pt}}
\multiput(714.00,293.17)(9.800,-5.000){2}{\rule{0.530pt}{0.400pt}}
\multiput(726.00,287.93)(1.378,-0.477){7}{\rule{1.140pt}{0.115pt}}
\multiput(726.00,288.17)(10.634,-5.000){2}{\rule{0.570pt}{0.400pt}}
\multiput(739.00,282.93)(1.267,-0.477){7}{\rule{1.060pt}{0.115pt}}
\multiput(739.00,283.17)(9.800,-5.000){2}{\rule{0.530pt}{0.400pt}}
\multiput(751.00,277.94)(1.797,-0.468){5}{\rule{1.400pt}{0.113pt}}
\multiput(751.00,278.17)(10.094,-4.000){2}{\rule{0.700pt}{0.400pt}}
\multiput(764.00,273.93)(1.267,-0.477){7}{\rule{1.060pt}{0.115pt}}
\multiput(764.00,274.17)(9.800,-5.000){2}{\rule{0.530pt}{0.400pt}}
\multiput(776.00,268.94)(1.797,-0.468){5}{\rule{1.400pt}{0.113pt}}
\multiput(776.00,269.17)(10.094,-4.000){2}{\rule{0.700pt}{0.400pt}}
\multiput(789.00,264.93)(1.267,-0.477){7}{\rule{1.060pt}{0.115pt}}
\multiput(789.00,265.17)(9.800,-5.000){2}{\rule{0.530pt}{0.400pt}}
\multiput(801.00,259.94)(1.797,-0.468){5}{\rule{1.400pt}{0.113pt}}
\multiput(801.00,260.17)(10.094,-4.000){2}{\rule{0.700pt}{0.400pt}}
\multiput(814.00,255.94)(1.651,-0.468){5}{\rule{1.300pt}{0.113pt}}
\multiput(814.00,256.17)(9.302,-4.000){2}{\rule{0.650pt}{0.400pt}}
\multiput(826.00,251.94)(1.797,-0.468){5}{\rule{1.400pt}{0.113pt}}
\multiput(826.00,252.17)(10.094,-4.000){2}{\rule{0.700pt}{0.400pt}}
\multiput(839.00,247.94)(1.651,-0.468){5}{\rule{1.300pt}{0.113pt}}
\multiput(839.00,248.17)(9.302,-4.000){2}{\rule{0.650pt}{0.400pt}}
\multiput(851.00,243.95)(2.695,-0.447){3}{\rule{1.833pt}{0.108pt}}
\multiput(851.00,244.17)(9.195,-3.000){2}{\rule{0.917pt}{0.400pt}}
\multiput(864.00,240.94)(1.651,-0.468){5}{\rule{1.300pt}{0.113pt}}
\multiput(864.00,241.17)(9.302,-4.000){2}{\rule{0.650pt}{0.400pt}}
\multiput(876.00,236.94)(1.797,-0.468){5}{\rule{1.400pt}{0.113pt}}
\multiput(876.00,237.17)(10.094,-4.000){2}{\rule{0.700pt}{0.400pt}}
\multiput(889.00,232.95)(2.472,-0.447){3}{\rule{1.700pt}{0.108pt}}
\multiput(889.00,233.17)(8.472,-3.000){2}{\rule{0.850pt}{0.400pt}}
\multiput(901.00,229.95)(2.695,-0.447){3}{\rule{1.833pt}{0.108pt}}
\multiput(901.00,230.17)(9.195,-3.000){2}{\rule{0.917pt}{0.400pt}}
\multiput(914.00,226.94)(1.651,-0.468){5}{\rule{1.300pt}{0.113pt}}
\multiput(914.00,227.17)(9.302,-4.000){2}{\rule{0.650pt}{0.400pt}}
\multiput(926.00,222.95)(2.695,-0.447){3}{\rule{1.833pt}{0.108pt}}
\multiput(926.00,223.17)(9.195,-3.000){2}{\rule{0.917pt}{0.400pt}}
\multiput(939.00,219.95)(2.472,-0.447){3}{\rule{1.700pt}{0.108pt}}
\multiput(939.00,220.17)(8.472,-3.000){2}{\rule{0.850pt}{0.400pt}}
\multiput(951.00,216.95)(2.695,-0.447){3}{\rule{1.833pt}{0.108pt}}
\multiput(951.00,217.17)(9.195,-3.000){2}{\rule{0.917pt}{0.400pt}}
\multiput(964.00,213.95)(2.472,-0.447){3}{\rule{1.700pt}{0.108pt}}
\multiput(964.00,214.17)(8.472,-3.000){2}{\rule{0.850pt}{0.400pt}}
\multiput(976.00,210.95)(2.695,-0.447){3}{\rule{1.833pt}{0.108pt}}
\multiput(976.00,211.17)(9.195,-3.000){2}{\rule{0.917pt}{0.400pt}}
\multiput(989.00,207.95)(2.472,-0.447){3}{\rule{1.700pt}{0.108pt}}
\multiput(989.00,208.17)(8.472,-3.000){2}{\rule{0.850pt}{0.400pt}}
\put(1001,204.17){\rule{2.700pt}{0.400pt}}
\multiput(1001.00,205.17)(7.396,-2.000){2}{\rule{1.350pt}{0.400pt}}
\multiput(1014.00,202.95)(2.472,-0.447){3}{\rule{1.700pt}{0.108pt}}
\multiput(1014.00,203.17)(8.472,-3.000){2}{\rule{0.850pt}{0.400pt}}
\multiput(1026.00,199.95)(2.695,-0.447){3}{\rule{1.833pt}{0.108pt}}
\multiput(1026.00,200.17)(9.195,-3.000){2}{\rule{0.917pt}{0.400pt}}
\put(1039,196.17){\rule{2.500pt}{0.400pt}}
\multiput(1039.00,197.17)(6.811,-2.000){2}{\rule{1.250pt}{0.400pt}}
\multiput(1051.00,194.95)(2.695,-0.447){3}{\rule{1.833pt}{0.108pt}}
\multiput(1051.00,195.17)(9.195,-3.000){2}{\rule{0.917pt}{0.400pt}}
\put(1064,191.17){\rule{2.500pt}{0.400pt}}
\multiput(1064.00,192.17)(6.811,-2.000){2}{\rule{1.250pt}{0.400pt}}
\put(1076,189.17){\rule{2.700pt}{0.400pt}}
\multiput(1076.00,190.17)(7.396,-2.000){2}{\rule{1.350pt}{0.400pt}}
\put(1089,187.17){\rule{2.500pt}{0.400pt}}
\multiput(1089.00,188.17)(6.811,-2.000){2}{\rule{1.250pt}{0.400pt}}
\multiput(1101.00,185.95)(2.695,-0.447){3}{\rule{1.833pt}{0.108pt}}
\multiput(1101.00,186.17)(9.195,-3.000){2}{\rule{0.917pt}{0.400pt}}
\put(1114,182.17){\rule{2.500pt}{0.400pt}}
\multiput(1114.00,183.17)(6.811,-2.000){2}{\rule{1.250pt}{0.400pt}}
\put(1126,180.17){\rule{2.700pt}{0.400pt}}
\multiput(1126.00,181.17)(7.396,-2.000){2}{\rule{1.350pt}{0.400pt}}
\put(1139,178.17){\rule{2.500pt}{0.400pt}}
\multiput(1139.00,179.17)(6.811,-2.000){2}{\rule{1.250pt}{0.400pt}}
\put(1151,176.17){\rule{2.700pt}{0.400pt}}
\multiput(1151.00,177.17)(7.396,-2.000){2}{\rule{1.350pt}{0.400pt}}
\put(1164,174.67){\rule{2.891pt}{0.400pt}}
\multiput(1164.00,175.17)(6.000,-1.000){2}{\rule{1.445pt}{0.400pt}}
\put(1176,173.17){\rule{2.700pt}{0.400pt}}
\multiput(1176.00,174.17)(7.396,-2.000){2}{\rule{1.350pt}{0.400pt}}
\put(1189,171.17){\rule{2.500pt}{0.400pt}}
\multiput(1189.00,172.17)(6.811,-2.000){2}{\rule{1.250pt}{0.400pt}}
\put(1201,169.67){\rule{3.132pt}{0.400pt}}
\multiput(1201.00,170.17)(6.500,-1.000){2}{\rule{1.566pt}{0.400pt}}
\put(1214,168.17){\rule{2.500pt}{0.400pt}}
\multiput(1214.00,169.17)(6.811,-2.000){2}{\rule{1.250pt}{0.400pt}}
\put(1226,166.17){\rule{2.700pt}{0.400pt}}
\multiput(1226.00,167.17)(7.396,-2.000){2}{\rule{1.350pt}{0.400pt}}
\put(1239,164.67){\rule{2.891pt}{0.400pt}}
\multiput(1239.00,165.17)(6.000,-1.000){2}{\rule{1.445pt}{0.400pt}}
\put(1251,163.17){\rule{2.700pt}{0.400pt}}
\multiput(1251.00,164.17)(7.396,-2.000){2}{\rule{1.350pt}{0.400pt}}
\put(1264,161.67){\rule{2.891pt}{0.400pt}}
\multiput(1264.00,162.17)(6.000,-1.000){2}{\rule{1.445pt}{0.400pt}}
\put(1276,160.67){\rule{3.132pt}{0.400pt}}
\multiput(1276.00,161.17)(6.500,-1.000){2}{\rule{1.566pt}{0.400pt}}
\put(1289,159.17){\rule{2.500pt}{0.400pt}}
\multiput(1289.00,160.17)(6.811,-2.000){2}{\rule{1.250pt}{0.400pt}}
\put(1301,157.67){\rule{3.132pt}{0.400pt}}
\multiput(1301.00,158.17)(6.500,-1.000){2}{\rule{1.566pt}{0.400pt}}
\put(1314,156.67){\rule{2.891pt}{0.400pt}}
\multiput(1314.00,157.17)(6.000,-1.000){2}{\rule{1.445pt}{0.400pt}}
\put(1326,155.67){\rule{3.132pt}{0.400pt}}
\multiput(1326.00,156.17)(6.500,-1.000){2}{\rule{1.566pt}{0.400pt}}
\put(1339,154.67){\rule{2.891pt}{0.400pt}}
\multiput(1339.00,155.17)(6.000,-1.000){2}{\rule{1.445pt}{0.400pt}}
\put(1351,153.67){\rule{3.132pt}{0.400pt}}
\multiput(1351.00,154.17)(6.500,-1.000){2}{\rule{1.566pt}{0.400pt}}
\put(1364,152.67){\rule{2.891pt}{0.400pt}}
\multiput(1364.00,153.17)(6.000,-1.000){2}{\rule{1.445pt}{0.400pt}}
\put(1376,151.67){\rule{3.132pt}{0.400pt}}
\multiput(1376.00,152.17)(6.500,-1.000){2}{\rule{1.566pt}{0.400pt}}
\put(1389,150.67){\rule{2.891pt}{0.400pt}}
\multiput(1389.00,151.17)(6.000,-1.000){2}{\rule{1.445pt}{0.400pt}}
\put(1401,149.67){\rule{3.132pt}{0.400pt}}
\multiput(1401.00,150.17)(6.500,-1.000){2}{\rule{1.566pt}{0.400pt}}
\put(1426,148.67){\rule{3.132pt}{0.400pt}}
\multiput(1426.00,149.17)(6.500,-1.000){2}{\rule{1.566pt}{0.400pt}}
\put(1414.0,150.0){\rule[-0.200pt]{2.891pt}{0.400pt}}
\end{picture}

%% file: nvsk1.tex
\setlength{\unitlength}{0.240900pt}
\ifx\plotpoint\undefined\newsavebox{\plotpoint}\fi
\sbox{\plotpoint}{\rule[-0.200pt]{0.400pt}{0.400pt}}%
\begin{picture}(1500,900)(0,0)
\font\gnuplot=cmr10 at 10pt
\gnuplot
\sbox{\plotpoint}{\rule[-0.200pt]{0.400pt}{0.400pt}}%
\put(181.0,82.0){\rule[-0.200pt]{4.818pt}{0.400pt}}
\put(161,82){\makebox(0,0)[r]{0}}
\put(1419.0,82.0){\rule[-0.200pt]{4.818pt}{0.400pt}}
\put(181.0,238.0){\rule[-0.200pt]{4.818pt}{0.400pt}}
\put(161,238){\makebox(0,0)[r]{0.01}}
\put(1419.0,238.0){\rule[-0.200pt]{4.818pt}{0.400pt}}
\put(181.0,393.0){\rule[-0.200pt]{4.818pt}{0.400pt}}
\put(161,393){\makebox(0,0)[r]{0.02}}
\put(1419.0,393.0){\rule[-0.200pt]{4.818pt}{0.400pt}}
\put(181.0,549.0){\rule[-0.200pt]{4.818pt}{0.400pt}}
\put(161,549){\makebox(0,0)[r]{0.03}}
\put(1419.0,549.0){\rule[-0.200pt]{4.818pt}{0.400pt}}
\put(181.0,704.0){\rule[-0.200pt]{4.818pt}{0.400pt}}
\put(161,704){\makebox(0,0)[r]{0.04}}
\put(1419.0,704.0){\rule[-0.200pt]{4.818pt}{0.400pt}}
\put(181.0,860.0){\rule[-0.200pt]{4.818pt}{0.400pt}}
\put(161,860){\makebox(0,0)[r]{0.05}}
\put(1419.0,860.0){\rule[-0.200pt]{4.818pt}{0.400pt}}
\put(181.0,82.0){\rule[-0.200pt]{0.400pt}{4.818pt}}
\put(181,41){\makebox(0,0){0}}
\put(181.0,840.0){\rule[-0.200pt]{0.400pt}{4.818pt}}
\put(433.0,82.0){\rule[-0.200pt]{0.400pt}{4.818pt}}
\put(433,41){\makebox(0,0){0.2}}
\put(433.0,840.0){\rule[-0.200pt]{0.400pt}{4.818pt}}
\put(684.0,82.0){\rule[-0.200pt]{0.400pt}{4.818pt}}
\put(684,41){\makebox(0,0){0.4}}
\put(684.0,840.0){\rule[-0.200pt]{0.400pt}{4.818pt}}
\put(936.0,82.0){\rule[-0.200pt]{0.400pt}{4.818pt}}
\put(936,41){\makebox(0,0){0.6}}
\put(936.0,840.0){\rule[-0.200pt]{0.400pt}{4.818pt}}
\put(1187.0,82.0){\rule[-0.200pt]{0.400pt}{4.818pt}}
\put(1187,41){\makebox(0,0){0.8}}
\put(1187.0,840.0){\rule[-0.200pt]{0.400pt}{4.818pt}}
\put(1439.0,82.0){\rule[-0.200pt]{0.400pt}{4.818pt}}
\put(1439,41){\makebox(0,0){1}}
\put(1439.0,840.0){\rule[-0.200pt]{0.400pt}{4.818pt}}
\put(181.0,82.0){\rule[-0.200pt]{303.052pt}{0.400pt}}
\put(1439.0,82.0){\rule[-0.200pt]{0.400pt}{187.420pt}}
\put(181.0,860.0){\rule[-0.200pt]{303.052pt}{0.400pt}}
\put(40,471){\makebox(0,0){\Large $n(k_\perp)$}}
\put(810,782){\makebox(0,0){\Huge b}}
\put(181.0,82.0){\rule[-0.200pt]{0.400pt}{187.420pt}}
\put(208.0,671.0){\rule[-0.200pt]{0.400pt}{8.913pt}}
\put(198.0,671.0){\rule[-0.200pt]{4.818pt}{0.400pt}}
\put(198.0,708.0){\rule[-0.200pt]{4.818pt}{0.400pt}}
\put(261.0,660.0){\rule[-0.200pt]{0.400pt}{6.263pt}}
\put(251.0,660.0){\rule[-0.200pt]{4.818pt}{0.400pt}}
\put(251.0,686.0){\rule[-0.200pt]{4.818pt}{0.400pt}}
\put(314.0,605.0){\rule[-0.200pt]{0.400pt}{6.263pt}}
\put(304.0,605.0){\rule[-0.200pt]{4.818pt}{0.400pt}}
\put(304.0,631.0){\rule[-0.200pt]{4.818pt}{0.400pt}}
\put(367.0,470.0){\rule[-0.200pt]{0.400pt}{4.336pt}}
\put(357.0,470.0){\rule[-0.200pt]{4.818pt}{0.400pt}}
\put(357.0,488.0){\rule[-0.200pt]{4.818pt}{0.400pt}}
\put(420.0,378.0){\rule[-0.200pt]{0.400pt}{3.854pt}}
\put(410.0,378.0){\rule[-0.200pt]{4.818pt}{0.400pt}}
\put(410.0,394.0){\rule[-0.200pt]{4.818pt}{0.400pt}}
\put(474.0,309.0){\rule[-0.200pt]{0.400pt}{2.891pt}}
\put(464.0,309.0){\rule[-0.200pt]{4.818pt}{0.400pt}}
\put(464.0,321.0){\rule[-0.200pt]{4.818pt}{0.400pt}}
\put(527.0,283.0){\rule[-0.200pt]{0.400pt}{2.650pt}}
\put(517.0,283.0){\rule[-0.200pt]{4.818pt}{0.400pt}}
\put(517.0,294.0){\rule[-0.200pt]{4.818pt}{0.400pt}}
\put(580.0,239.0){\rule[-0.200pt]{0.400pt}{1.927pt}}
\put(570.0,239.0){\rule[-0.200pt]{4.818pt}{0.400pt}}
\put(570.0,247.0){\rule[-0.200pt]{4.818pt}{0.400pt}}
\put(633.0,218.0){\rule[-0.200pt]{0.400pt}{1.445pt}}
\put(623.0,218.0){\rule[-0.200pt]{4.818pt}{0.400pt}}
\put(623.0,224.0){\rule[-0.200pt]{4.818pt}{0.400pt}}
\put(686.0,199.0){\rule[-0.200pt]{0.400pt}{1.445pt}}
\put(676.0,199.0){\rule[-0.200pt]{4.818pt}{0.400pt}}
\put(676.0,205.0){\rule[-0.200pt]{4.818pt}{0.400pt}}
\put(740.0,185.0){\rule[-0.200pt]{0.400pt}{1.204pt}}
\put(730.0,185.0){\rule[-0.200pt]{4.818pt}{0.400pt}}
\put(730.0,190.0){\rule[-0.200pt]{4.818pt}{0.400pt}}
\put(793.0,176.0){\rule[-0.200pt]{0.400pt}{0.964pt}}
\put(783.0,176.0){\rule[-0.200pt]{4.818pt}{0.400pt}}
\put(783.0,180.0){\rule[-0.200pt]{4.818pt}{0.400pt}}
\put(846.0,164.0){\rule[-0.200pt]{0.400pt}{1.204pt}}
\put(836.0,164.0){\rule[-0.200pt]{4.818pt}{0.400pt}}
\put(836.0,169.0){\rule[-0.200pt]{4.818pt}{0.400pt}}
\put(899.0,159.0){\rule[-0.200pt]{0.400pt}{0.964pt}}
\put(889.0,159.0){\rule[-0.200pt]{4.818pt}{0.400pt}}
\put(889.0,163.0){\rule[-0.200pt]{4.818pt}{0.400pt}}
\put(952.0,151.0){\rule[-0.200pt]{0.400pt}{0.723pt}}
\put(942.0,151.0){\rule[-0.200pt]{4.818pt}{0.400pt}}
\put(942.0,154.0){\rule[-0.200pt]{4.818pt}{0.400pt}}
\put(1006.0,146.0){\rule[-0.200pt]{0.400pt}{0.964pt}}
\put(996.0,146.0){\rule[-0.200pt]{4.818pt}{0.400pt}}
\put(996.0,150.0){\rule[-0.200pt]{4.818pt}{0.400pt}}
\put(1059.0,139.0){\rule[-0.200pt]{0.400pt}{0.723pt}}
\put(1049.0,139.0){\rule[-0.200pt]{4.818pt}{0.400pt}}
\put(1049.0,142.0){\rule[-0.200pt]{4.818pt}{0.400pt}}
\put(1112.0,136.0){\rule[-0.200pt]{0.400pt}{0.723pt}}
\put(1102.0,136.0){\rule[-0.200pt]{4.818pt}{0.400pt}}
\put(1102.0,139.0){\rule[-0.200pt]{4.818pt}{0.400pt}}
\put(1165.0,130.0){\rule[-0.200pt]{0.400pt}{0.482pt}}
\put(1155.0,130.0){\rule[-0.200pt]{4.818pt}{0.400pt}}
\put(1155.0,132.0){\rule[-0.200pt]{4.818pt}{0.400pt}}
\put(1218.0,127.0){\rule[-0.200pt]{0.400pt}{0.482pt}}
\put(1208.0,127.0){\rule[-0.200pt]{4.818pt}{0.400pt}}
\put(1208.0,129.0){\rule[-0.200pt]{4.818pt}{0.400pt}}
\put(1272.0,124.0){\rule[-0.200pt]{0.400pt}{0.482pt}}
\put(1262.0,124.0){\rule[-0.200pt]{4.818pt}{0.400pt}}
\put(1262.0,126.0){\rule[-0.200pt]{4.818pt}{0.400pt}}
\put(1325.0,120.0){\rule[-0.200pt]{0.400pt}{0.482pt}}
\put(1315.0,120.0){\rule[-0.200pt]{4.818pt}{0.400pt}}
\put(1315.0,122.0){\rule[-0.200pt]{4.818pt}{0.400pt}}
\put(1378.0,118.0){\rule[-0.200pt]{0.400pt}{0.482pt}}
\put(1368.0,118.0){\rule[-0.200pt]{4.818pt}{0.400pt}}
\put(1368.0,120.0){\rule[-0.200pt]{4.818pt}{0.400pt}}
\put(1431.0,116.0){\rule[-0.200pt]{0.400pt}{0.482pt}}
\put(1421.0,116.0){\rule[-0.200pt]{4.818pt}{0.400pt}}
\put(208,689){\raisebox{-.8pt}{\makebox(0,0){$\Diamond$}}}
\put(261,673){\raisebox{-.8pt}{\makebox(0,0){$\Diamond$}}}
\put(314,618){\raisebox{-.8pt}{\makebox(0,0){$\Diamond$}}}
\put(367,479){\raisebox{-.8pt}{\makebox(0,0){$\Diamond$}}}
\put(420,386){\raisebox{-.8pt}{\makebox(0,0){$\Diamond$}}}
\put(474,315){\raisebox{-.8pt}{\makebox(0,0){$\Diamond$}}}
\put(527,289){\raisebox{-.8pt}{\makebox(0,0){$\Diamond$}}}
\put(580,243){\raisebox{-.8pt}{\makebox(0,0){$\Diamond$}}}
\put(633,221){\raisebox{-.8pt}{\makebox(0,0){$\Diamond$}}}
\put(686,202){\raisebox{-.8pt}{\makebox(0,0){$\Diamond$}}}
\put(740,187){\raisebox{-.8pt}{\makebox(0,0){$\Diamond$}}}
\put(793,178){\raisebox{-.8pt}{\makebox(0,0){$\Diamond$}}}
\put(846,167){\raisebox{-.8pt}{\makebox(0,0){$\Diamond$}}}
\put(899,161){\raisebox{-.8pt}{\makebox(0,0){$\Diamond$}}}
\put(952,152){\raisebox{-.8pt}{\makebox(0,0){$\Diamond$}}}
\put(1006,148){\raisebox{-.8pt}{\makebox(0,0){$\Diamond$}}}
\put(1059,141){\raisebox{-.8pt}{\makebox(0,0){$\Diamond$}}}
\put(1112,137){\raisebox{-.8pt}{\makebox(0,0){$\Diamond$}}}
\put(1165,131){\raisebox{-.8pt}{\makebox(0,0){$\Diamond$}}}
\put(1218,128){\raisebox{-.8pt}{\makebox(0,0){$\Diamond$}}}
\put(1272,125){\raisebox{-.8pt}{\makebox(0,0){$\Diamond$}}}
\put(1325,121){\raisebox{-.8pt}{\makebox(0,0){$\Diamond$}}}
\put(1378,119){\raisebox{-.8pt}{\makebox(0,0){$\Diamond$}}}
\put(1431,117){\raisebox{-.8pt}{\makebox(0,0){$\Diamond$}}}
\put(1421.0,118.0){\rule[-0.200pt]{4.818pt}{0.400pt}}
\put(181.0,726.0){\rule[-0.200pt]{0.400pt}{15.899pt}}
\put(171.0,726.0){\rule[-0.200pt]{4.818pt}{0.400pt}}
\put(171.0,792.0){\rule[-0.200pt]{4.818pt}{0.400pt}}
\put(234.0,779.0){\rule[-0.200pt]{0.400pt}{8.672pt}}
\put(224.0,779.0){\rule[-0.200pt]{4.818pt}{0.400pt}}
\put(224.0,815.0){\rule[-0.200pt]{4.818pt}{0.400pt}}
\put(287.0,656.0){\rule[-0.200pt]{0.400pt}{5.300pt}}
\put(277.0,656.0){\rule[-0.200pt]{4.818pt}{0.400pt}}
\put(277.0,678.0){\rule[-0.200pt]{4.818pt}{0.400pt}}
\put(341.0,520.0){\rule[-0.200pt]{0.400pt}{4.336pt}}
\put(331.0,520.0){\rule[-0.200pt]{4.818pt}{0.400pt}}
\put(331.0,538.0){\rule[-0.200pt]{4.818pt}{0.400pt}}
\put(394.0,428.0){\rule[-0.200pt]{0.400pt}{4.336pt}}
\put(384.0,428.0){\rule[-0.200pt]{4.818pt}{0.400pt}}
\put(384.0,446.0){\rule[-0.200pt]{4.818pt}{0.400pt}}
\put(447.0,337.0){\rule[-0.200pt]{0.400pt}{2.650pt}}
\put(437.0,337.0){\rule[-0.200pt]{4.818pt}{0.400pt}}
\put(437.0,348.0){\rule[-0.200pt]{4.818pt}{0.400pt}}
\put(500.0,272.0){\rule[-0.200pt]{0.400pt}{2.168pt}}
\put(490.0,272.0){\rule[-0.200pt]{4.818pt}{0.400pt}}
\put(490.0,281.0){\rule[-0.200pt]{4.818pt}{0.400pt}}
\put(553.0,249.0){\rule[-0.200pt]{0.400pt}{1.927pt}}
\put(543.0,249.0){\rule[-0.200pt]{4.818pt}{0.400pt}}
\put(543.0,257.0){\rule[-0.200pt]{4.818pt}{0.400pt}}
\put(607.0,215.0){\rule[-0.200pt]{0.400pt}{1.445pt}}
\put(597.0,215.0){\rule[-0.200pt]{4.818pt}{0.400pt}}
\put(597.0,221.0){\rule[-0.200pt]{4.818pt}{0.400pt}}
\put(660.0,199.0){\rule[-0.200pt]{0.400pt}{1.445pt}}
\put(650.0,199.0){\rule[-0.200pt]{4.818pt}{0.400pt}}
\put(650.0,205.0){\rule[-0.200pt]{4.818pt}{0.400pt}}
\put(713.0,186.0){\rule[-0.200pt]{0.400pt}{1.445pt}}
\put(703.0,186.0){\rule[-0.200pt]{4.818pt}{0.400pt}}
\put(703.0,192.0){\rule[-0.200pt]{4.818pt}{0.400pt}}
\put(766.0,172.0){\rule[-0.200pt]{0.400pt}{1.204pt}}
\put(756.0,172.0){\rule[-0.200pt]{4.818pt}{0.400pt}}
\put(756.0,177.0){\rule[-0.200pt]{4.818pt}{0.400pt}}
\put(819.0,163.0){\rule[-0.200pt]{0.400pt}{0.723pt}}
\put(809.0,163.0){\rule[-0.200pt]{4.818pt}{0.400pt}}
\put(809.0,166.0){\rule[-0.200pt]{4.818pt}{0.400pt}}
\put(873.0,157.0){\rule[-0.200pt]{0.400pt}{0.964pt}}
\put(863.0,157.0){\rule[-0.200pt]{4.818pt}{0.400pt}}
\put(863.0,161.0){\rule[-0.200pt]{4.818pt}{0.400pt}}
\put(926.0,145.0){\rule[-0.200pt]{0.400pt}{0.723pt}}
\put(916.0,145.0){\rule[-0.200pt]{4.818pt}{0.400pt}}
\put(916.0,148.0){\rule[-0.200pt]{4.818pt}{0.400pt}}
\put(979.0,144.0){\rule[-0.200pt]{0.400pt}{0.482pt}}
\put(969.0,144.0){\rule[-0.200pt]{4.818pt}{0.400pt}}
\put(969.0,146.0){\rule[-0.200pt]{4.818pt}{0.400pt}}
\put(1032.0,138.0){\rule[-0.200pt]{0.400pt}{0.723pt}}
\put(1022.0,138.0){\rule[-0.200pt]{4.818pt}{0.400pt}}
\put(1022.0,141.0){\rule[-0.200pt]{4.818pt}{0.400pt}}
\put(1085.0,132.0){\rule[-0.200pt]{0.400pt}{0.723pt}}
\put(1075.0,132.0){\rule[-0.200pt]{4.818pt}{0.400pt}}
\put(1075.0,135.0){\rule[-0.200pt]{4.818pt}{0.400pt}}
\put(1139.0,128.0){\rule[-0.200pt]{0.400pt}{0.482pt}}
\put(1129.0,128.0){\rule[-0.200pt]{4.818pt}{0.400pt}}
\put(1129.0,130.0){\rule[-0.200pt]{4.818pt}{0.400pt}}
\put(1192.0,125.0){\rule[-0.200pt]{0.400pt}{0.482pt}}
\put(1182.0,125.0){\rule[-0.200pt]{4.818pt}{0.400pt}}
\put(1182.0,127.0){\rule[-0.200pt]{4.818pt}{0.400pt}}
\put(1245.0,121.0){\usebox{\plotpoint}}
\put(1235.0,121.0){\rule[-0.200pt]{4.818pt}{0.400pt}}
\put(1235.0,122.0){\rule[-0.200pt]{4.818pt}{0.400pt}}
\put(1298.0,120.0){\rule[-0.200pt]{0.400pt}{0.482pt}}
\put(1288.0,120.0){\rule[-0.200pt]{4.818pt}{0.400pt}}
\put(1288.0,122.0){\rule[-0.200pt]{4.818pt}{0.400pt}}
\put(1351.0,116.0){\usebox{\plotpoint}}
\put(1341.0,116.0){\rule[-0.200pt]{4.818pt}{0.400pt}}
\put(1341.0,117.0){\rule[-0.200pt]{4.818pt}{0.400pt}}
\put(1405.0,114.0){\rule[-0.200pt]{0.400pt}{0.482pt}}
\put(1395.0,114.0){\rule[-0.200pt]{4.818pt}{0.400pt}}
\put(181,759){\makebox(0,0){$+$}}
\put(234,797){\makebox(0,0){$+$}}
\put(287,667){\makebox(0,0){$+$}}
\put(341,529){\makebox(0,0){$+$}}
\put(394,437){\makebox(0,0){$+$}}
\put(447,343){\makebox(0,0){$+$}}
\put(500,276){\makebox(0,0){$+$}}
\put(553,253){\makebox(0,0){$+$}}
\put(607,218){\makebox(0,0){$+$}}
\put(660,202){\makebox(0,0){$+$}}
\put(713,189){\makebox(0,0){$+$}}
\put(766,175){\makebox(0,0){$+$}}
\put(819,165){\makebox(0,0){$+$}}
\put(873,159){\makebox(0,0){$+$}}
\put(926,147){\makebox(0,0){$+$}}
\put(979,145){\makebox(0,0){$+$}}
\put(1032,139){\makebox(0,0){$+$}}
\put(1085,134){\makebox(0,0){$+$}}
\put(1139,129){\makebox(0,0){$+$}}
\put(1192,126){\makebox(0,0){$+$}}
\put(1245,121){\makebox(0,0){$+$}}
\put(1298,121){\makebox(0,0){$+$}}
\put(1351,117){\makebox(0,0){$+$}}
\put(1405,115){\makebox(0,0){$+$}}
\put(1395.0,116.0){\rule[-0.200pt]{4.818pt}{0.400pt}}
\end{picture}

%% file: nvsmul.tex
\setlength{\unitlength}{0.240900pt}
\ifx\plotpoint\undefined\newsavebox{\plotpoint}\fi
\sbox{\plotpoint}{\rule[-0.200pt]{0.400pt}{0.400pt}}%
\begin{picture}(1500,900)(0,0)
\font\gnuplot=cmr10 at 10pt
\gnuplot
\sbox{\plotpoint}{\rule[-0.200pt]{0.400pt}{0.400pt}}%
\put(181.0,123.0){\rule[-0.200pt]{4.818pt}{0.400pt}}
\put(161,123){\makebox(0,0)[r]{0.11}}
\put(1419.0,123.0){\rule[-0.200pt]{4.818pt}{0.400pt}}
\put(181.0,270.0){\rule[-0.200pt]{4.818pt}{0.400pt}}
\put(161,270){\makebox(0,0)[r]{0.12}}
\put(1419.0,270.0){\rule[-0.200pt]{4.818pt}{0.400pt}}
\put(181.0,418.0){\rule[-0.200pt]{4.818pt}{0.400pt}}
\put(161,418){\makebox(0,0)[r]{0.13}}
\put(1419.0,418.0){\rule[-0.200pt]{4.818pt}{0.400pt}}
\put(181.0,565.0){\rule[-0.200pt]{4.818pt}{0.400pt}}
\put(161,565){\makebox(0,0)[r]{0.14}}
\put(1419.0,565.0){\rule[-0.200pt]{4.818pt}{0.400pt}}
\put(181.0,713.0){\rule[-0.200pt]{4.818pt}{0.400pt}}
\put(161,713){\makebox(0,0)[r]{0.15}}
\put(1419.0,713.0){\rule[-0.200pt]{4.818pt}{0.400pt}}
\put(181.0,860.0){\rule[-0.200pt]{4.818pt}{0.400pt}}
\put(161,860){\makebox(0,0)[r]{0.16}}
\put(1419.0,860.0){\rule[-0.200pt]{4.818pt}{0.400pt}}
\put(443.0,123.0){\rule[-0.200pt]{0.400pt}{4.818pt}}
\put(443,82){\makebox(0,0){50}}
\put(443.0,840.0){\rule[-0.200pt]{0.400pt}{4.818pt}}
\put(798.0,123.0){\rule[-0.200pt]{0.400pt}{4.818pt}}
\put(798,82){\makebox(0,0){100}}
\put(798.0,840.0){\rule[-0.200pt]{0.400pt}{4.818pt}}
\put(1152.0,123.0){\rule[-0.200pt]{0.400pt}{4.818pt}}
\put(1152,82){\makebox(0,0){200}}
\put(1152.0,840.0){\rule[-0.200pt]{0.400pt}{4.818pt}}
\put(1360.0,123.0){\rule[-0.200pt]{0.400pt}{4.818pt}}
\put(1360,82){\makebox(0,0){300}}
\put(1360.0,840.0){\rule[-0.200pt]{0.400pt}{4.818pt}}
\put(181.0,123.0){\rule[-0.200pt]{303.052pt}{0.400pt}}
\put(1439.0,123.0){\rule[-0.200pt]{0.400pt}{177.543pt}}
\put(181.0,860.0){\rule[-0.200pt]{303.052pt}{0.400pt}}
\put(40,491){\makebox(0,0){\Large $f_N$}}
\put(810,21){\makebox(0,0){\Large $g^2\mu L$}}
\put(181.0,123.0){\rule[-0.200pt]{0.400pt}{177.543pt}}
\put(265.0,341.0){\rule[-0.200pt]{0.400pt}{13.009pt}}
\put(255.0,341.0){\rule[-0.200pt]{4.818pt}{0.400pt}}
\put(255.0,395.0){\rule[-0.200pt]{4.818pt}{0.400pt}}
\put(620.0,323.0){\rule[-0.200pt]{0.400pt}{10.840pt}}
\put(610.0,323.0){\rule[-0.200pt]{4.818pt}{0.400pt}}
\put(610.0,368.0){\rule[-0.200pt]{4.818pt}{0.400pt}}
\put(828.0,476.0){\rule[-0.200pt]{0.400pt}{7.950pt}}
\put(818.0,476.0){\rule[-0.200pt]{4.818pt}{0.400pt}}
\put(818.0,509.0){\rule[-0.200pt]{4.818pt}{0.400pt}}
\put(1000.0,574.0){\rule[-0.200pt]{0.400pt}{7.950pt}}
\put(990.0,574.0){\rule[-0.200pt]{4.818pt}{0.400pt}}
\put(990.0,607.0){\rule[-0.200pt]{4.818pt}{0.400pt}}
\put(1183.0,620.0){\rule[-0.200pt]{0.400pt}{9.154pt}}
\put(1173.0,620.0){\rule[-0.200pt]{4.818pt}{0.400pt}}
\put(1173.0,658.0){\rule[-0.200pt]{4.818pt}{0.400pt}}
\put(1355.0,741.0){\rule[-0.200pt]{0.400pt}{8.913pt}}
\put(1345.0,741.0){\rule[-0.200pt]{4.818pt}{0.400pt}}
\put(265,368){\raisebox{-.8pt}{\makebox(0,0){$\Diamond$}}}
\put(620,346){\raisebox{-.8pt}{\makebox(0,0){$\Diamond$}}}
\put(828,493){\raisebox{-.8pt}{\makebox(0,0){$\Diamond$}}}
\put(1000,591){\raisebox{-.8pt}{\makebox(0,0){$\Diamond$}}}
\put(1183,639){\raisebox{-.8pt}{\makebox(0,0){$\Diamond$}}}
\put(1355,759){\raisebox{-.8pt}{\makebox(0,0){$\Diamond$}}}
\put(1345.0,778.0){\rule[-0.200pt]{4.818pt}{0.400pt}}
\put(265.0,192.0){\rule[-0.200pt]{0.400pt}{10.359pt}}
\put(255.0,192.0){\rule[-0.200pt]{4.818pt}{0.400pt}}
\put(255.0,235.0){\rule[-0.200pt]{4.818pt}{0.400pt}}
\put(620.0,236.0){\rule[-0.200pt]{0.400pt}{10.118pt}}
\put(610.0,236.0){\rule[-0.200pt]{4.818pt}{0.400pt}}
\put(610.0,278.0){\rule[-0.200pt]{4.818pt}{0.400pt}}
\put(828.0,359.0){\rule[-0.200pt]{0.400pt}{7.950pt}}
\put(818.0,359.0){\rule[-0.200pt]{4.818pt}{0.400pt}}
\put(818.0,392.0){\rule[-0.200pt]{4.818pt}{0.400pt}}
\put(1000.0,516.0){\rule[-0.200pt]{0.400pt}{8.191pt}}
\put(990.0,516.0){\rule[-0.200pt]{4.818pt}{0.400pt}}
\put(990.0,550.0){\rule[-0.200pt]{4.818pt}{0.400pt}}
\put(1183.0,640.0){\rule[-0.200pt]{0.400pt}{9.636pt}}
\put(1173.0,640.0){\rule[-0.200pt]{4.818pt}{0.400pt}}
\put(1173.0,680.0){\rule[-0.200pt]{4.818pt}{0.400pt}}
\put(1355.0,702.0){\rule[-0.200pt]{0.400pt}{8.672pt}}
\put(1345.0,702.0){\rule[-0.200pt]{4.818pt}{0.400pt}}
\put(265,213){\makebox(0,0){$+$}}
\put(620,257){\makebox(0,0){$+$}}
\put(828,375){\makebox(0,0){$+$}}
\put(1000,533){\makebox(0,0){$+$}}
\put(1183,660){\makebox(0,0){$+$}}
\put(1355,720){\makebox(0,0){$+$}}
\put(1345.0,738.0){\rule[-0.200pt]{4.818pt}{0.400pt}}
\end{picture}

%% file: dr.tex
\setlength{\unitlength}{0.240900pt}
\ifx\plotpoint\undefined\newsavebox{\plotpoint}\fi
\sbox{\plotpoint}{\rule[-0.200pt]{0.400pt}{0.400pt}}%
\begin{picture}(1500,900)(0,0)
\font\gnuplot=cmr10 at 10pt
\gnuplot
\sbox{\plotpoint}{\rule[-0.200pt]{0.400pt}{0.400pt}}%
\put(161.0,123.0){\rule[-0.200pt]{4.818pt}{0.400pt}}
\put(141,123){\makebox(0,0)[r]{0}}
\put(1419.0,123.0){\rule[-0.200pt]{4.818pt}{0.400pt}}
\put(161.0,270.0){\rule[-0.200pt]{4.818pt}{0.400pt}}
\put(141,270){\makebox(0,0)[r]{0.1}}
\put(1419.0,270.0){\rule[-0.200pt]{4.818pt}{0.400pt}}
\put(161.0,418.0){\rule[-0.200pt]{4.818pt}{0.400pt}}
\put(141,418){\makebox(0,0)[r]{0.2}}
\put(1419.0,418.0){\rule[-0.200pt]{4.818pt}{0.400pt}}
\put(161.0,565.0){\rule[-0.200pt]{4.818pt}{0.400pt}}
\put(141,565){\makebox(0,0)[r]{0.3}}
\put(1419.0,565.0){\rule[-0.200pt]{4.818pt}{0.400pt}}
\put(161.0,713.0){\rule[-0.200pt]{4.818pt}{0.400pt}}
\put(141,713){\makebox(0,0)[r]{0.4}}
\put(1419.0,713.0){\rule[-0.200pt]{4.818pt}{0.400pt}}
\put(161.0,860.0){\rule[-0.200pt]{4.818pt}{0.400pt}}
\put(141,860){\makebox(0,0)[r]{0.5}}
\put(1419.0,860.0){\rule[-0.200pt]{4.818pt}{0.400pt}}
\put(161.0,123.0){\rule[-0.200pt]{0.400pt}{4.818pt}}
\put(161,82){\makebox(0,0){0}}
\put(161.0,840.0){\rule[-0.200pt]{0.400pt}{4.818pt}}
\put(417.0,123.0){\rule[-0.200pt]{0.400pt}{4.818pt}}
\put(417,82){\makebox(0,0){0.1}}
\put(417.0,840.0){\rule[-0.200pt]{0.400pt}{4.818pt}}
\put(672.0,123.0){\rule[-0.200pt]{0.400pt}{4.818pt}}
\put(672,82){\makebox(0,0){0.2}}
\put(672.0,840.0){\rule[-0.200pt]{0.400pt}{4.818pt}}
\put(928.0,123.0){\rule[-0.200pt]{0.400pt}{4.818pt}}
\put(928,82){\makebox(0,0){0.3}}
\put(928.0,840.0){\rule[-0.200pt]{0.400pt}{4.818pt}}
\put(1183.0,123.0){\rule[-0.200pt]{0.400pt}{4.818pt}}
\put(1183,82){\makebox(0,0){0.4}}
\put(1183.0,840.0){\rule[-0.200pt]{0.400pt}{4.818pt}}
\put(1439.0,123.0){\rule[-0.200pt]{0.400pt}{4.818pt}}
\put(1439,82){\makebox(0,0){0.5}}
\put(1439.0,840.0){\rule[-0.200pt]{0.400pt}{4.818pt}}
\put(161.0,123.0){\rule[-0.200pt]{307.870pt}{0.400pt}}
\put(1439.0,123.0){\rule[-0.200pt]{0.400pt}{177.543pt}}
\put(161.0,860.0){\rule[-0.200pt]{307.870pt}{0.400pt}}
\put(40,491){\makebox(0,0){\Large ${\omega\over{g^2\mu}}$}}
\put(800,21){\makebox(0,0){\Large ${k_\perp\over{g^2\mu}}$}}
\put(161.0,123.0){\rule[-0.200pt]{0.400pt}{177.543pt}}
\put(161.0,403.0){\rule[-0.200pt]{0.400pt}{9.636pt}}
\put(151.0,403.0){\rule[-0.200pt]{4.818pt}{0.400pt}}
\put(151.0,443.0){\rule[-0.200pt]{4.818pt}{0.400pt}}
\put(388.0,388.0){\rule[-0.200pt]{0.400pt}{4.818pt}}
\put(378.0,388.0){\rule[-0.200pt]{4.818pt}{0.400pt}}
\put(378.0,408.0){\rule[-0.200pt]{4.818pt}{0.400pt}}
\put(615.0,388.0){\rule[-0.200pt]{0.400pt}{4.336pt}}
\put(605.0,388.0){\rule[-0.200pt]{4.818pt}{0.400pt}}
\put(605.0,406.0){\rule[-0.200pt]{4.818pt}{0.400pt}}
\put(842.0,480.0){\rule[-0.200pt]{0.400pt}{6.745pt}}
\put(832.0,480.0){\rule[-0.200pt]{4.818pt}{0.400pt}}
\put(832.0,508.0){\rule[-0.200pt]{4.818pt}{0.400pt}}
\put(1069.0,581.0){\rule[-0.200pt]{0.400pt}{7.950pt}}
\put(1059.0,581.0){\rule[-0.200pt]{4.818pt}{0.400pt}}
\put(1059.0,614.0){\rule[-0.200pt]{4.818pt}{0.400pt}}
\put(1296.0,737.0){\rule[-0.200pt]{0.400pt}{9.877pt}}
\put(1286.0,737.0){\rule[-0.200pt]{4.818pt}{0.400pt}}
\put(161,423){\raisebox{-.8pt}{\makebox(0,0){$\Diamond$}}}
\put(388,398){\raisebox{-.8pt}{\makebox(0,0){$\Diamond$}}}
\put(615,397){\raisebox{-.8pt}{\makebox(0,0){$\Diamond$}}}
\put(842,494){\raisebox{-.8pt}{\makebox(0,0){$\Diamond$}}}
\put(1069,597){\raisebox{-.8pt}{\makebox(0,0){$\Diamond$}}}
\put(1296,757){\raisebox{-.8pt}{\makebox(0,0){$\Diamond$}}}
\put(1286.0,778.0){\rule[-0.200pt]{4.818pt}{0.400pt}}
\put(161.0,244.0){\rule[-0.200pt]{0.400pt}{3.854pt}}
\put(151.0,244.0){\rule[-0.200pt]{4.818pt}{0.400pt}}
\put(151.0,260.0){\rule[-0.200pt]{4.818pt}{0.400pt}}
\put(269.0,251.0){\rule[-0.200pt]{0.400pt}{2.409pt}}
\put(259.0,251.0){\rule[-0.200pt]{4.818pt}{0.400pt}}
\put(259.0,261.0){\rule[-0.200pt]{4.818pt}{0.400pt}}
\put(377.0,269.0){\rule[-0.200pt]{0.400pt}{2.168pt}}
\put(367.0,269.0){\rule[-0.200pt]{4.818pt}{0.400pt}}
\put(367.0,278.0){\rule[-0.200pt]{4.818pt}{0.400pt}}
\put(485.0,313.0){\rule[-0.200pt]{0.400pt}{3.132pt}}
\put(475.0,313.0){\rule[-0.200pt]{4.818pt}{0.400pt}}
\put(475.0,326.0){\rule[-0.200pt]{4.818pt}{0.400pt}}
\put(593.0,360.0){\rule[-0.200pt]{0.400pt}{3.854pt}}
\put(583.0,360.0){\rule[-0.200pt]{4.818pt}{0.400pt}}
\put(583.0,376.0){\rule[-0.200pt]{4.818pt}{0.400pt}}
\put(701.0,418.0){\rule[-0.200pt]{0.400pt}{4.095pt}}
\put(691.0,418.0){\rule[-0.200pt]{4.818pt}{0.400pt}}
\put(691.0,435.0){\rule[-0.200pt]{4.818pt}{0.400pt}}
\put(810.0,469.0){\rule[-0.200pt]{0.400pt}{5.541pt}}
\put(800.0,469.0){\rule[-0.200pt]{4.818pt}{0.400pt}}
\put(800.0,492.0){\rule[-0.200pt]{4.818pt}{0.400pt}}
\put(918.0,535.0){\rule[-0.200pt]{0.400pt}{6.263pt}}
\put(908.0,535.0){\rule[-0.200pt]{4.818pt}{0.400pt}}
\put(908.0,561.0){\rule[-0.200pt]{4.818pt}{0.400pt}}
\put(1026.0,586.0){\rule[-0.200pt]{0.400pt}{8.191pt}}
\put(1016.0,586.0){\rule[-0.200pt]{4.818pt}{0.400pt}}
\put(1016.0,620.0){\rule[-0.200pt]{4.818pt}{0.400pt}}
\put(1134.0,639.0){\rule[-0.200pt]{0.400pt}{8.191pt}}
\put(1124.0,639.0){\rule[-0.200pt]{4.818pt}{0.400pt}}
\put(1124.0,673.0){\rule[-0.200pt]{4.818pt}{0.400pt}}
\put(1242.0,720.0){\rule[-0.200pt]{0.400pt}{10.118pt}}
\put(1232.0,720.0){\rule[-0.200pt]{4.818pt}{0.400pt}}
\put(1232.0,762.0){\rule[-0.200pt]{4.818pt}{0.400pt}}
\put(1350.0,777.0){\rule[-0.200pt]{0.400pt}{11.563pt}}
\put(1340.0,777.0){\rule[-0.200pt]{4.818pt}{0.400pt}}
\put(161,252){\makebox(0,0){$+$}}
\put(269,256){\makebox(0,0){$+$}}
\put(377,274){\makebox(0,0){$+$}}
\put(485,320){\makebox(0,0){$+$}}
\put(593,368){\makebox(0,0){$+$}}
\put(701,426){\makebox(0,0){$+$}}
\put(810,480){\makebox(0,0){$+$}}
\put(918,548){\makebox(0,0){$+$}}
\put(1026,603){\makebox(0,0){$+$}}
\put(1134,656){\makebox(0,0){$+$}}
\put(1242,741){\makebox(0,0){$+$}}
\put(1350,801){\makebox(0,0){$+$}}
\put(1340.0,825.0){\rule[-0.200pt]{4.818pt}{0.400pt}}
\put(161.0,290.0){\rule[-0.200pt]{0.400pt}{6.022pt}}
\put(151.0,290.0){\rule[-0.200pt]{4.818pt}{0.400pt}}
\put(151.0,315.0){\rule[-0.200pt]{4.818pt}{0.400pt}}
\put(215.0,288.0){\rule[-0.200pt]{0.400pt}{3.132pt}}
\put(205.0,288.0){\rule[-0.200pt]{4.818pt}{0.400pt}}
\put(205.0,301.0){\rule[-0.200pt]{4.818pt}{0.400pt}}
\put(269.0,285.0){\rule[-0.200pt]{0.400pt}{2.891pt}}
\put(259.0,285.0){\rule[-0.200pt]{4.818pt}{0.400pt}}
\put(259.0,297.0){\rule[-0.200pt]{4.818pt}{0.400pt}}
\put(323.0,280.0){\rule[-0.200pt]{0.400pt}{2.650pt}}
\put(313.0,280.0){\rule[-0.200pt]{4.818pt}{0.400pt}}
\put(313.0,291.0){\rule[-0.200pt]{4.818pt}{0.400pt}}
\put(377.0,290.0){\rule[-0.200pt]{0.400pt}{2.891pt}}
\put(367.0,290.0){\rule[-0.200pt]{4.818pt}{0.400pt}}
\put(367.0,302.0){\rule[-0.200pt]{4.818pt}{0.400pt}}
\put(431.0,305.0){\rule[-0.200pt]{0.400pt}{2.891pt}}
\put(421.0,305.0){\rule[-0.200pt]{4.818pt}{0.400pt}}
\put(421.0,317.0){\rule[-0.200pt]{4.818pt}{0.400pt}}
\put(485.0,321.0){\rule[-0.200pt]{0.400pt}{3.373pt}}
\put(475.0,321.0){\rule[-0.200pt]{4.818pt}{0.400pt}}
\put(475.0,335.0){\rule[-0.200pt]{4.818pt}{0.400pt}}
\put(539.0,336.0){\rule[-0.200pt]{0.400pt}{3.854pt}}
\put(529.0,336.0){\rule[-0.200pt]{4.818pt}{0.400pt}}
\put(529.0,352.0){\rule[-0.200pt]{4.818pt}{0.400pt}}
\put(593.0,357.0){\rule[-0.200pt]{0.400pt}{3.854pt}}
\put(583.0,357.0){\rule[-0.200pt]{4.818pt}{0.400pt}}
\put(583.0,373.0){\rule[-0.200pt]{4.818pt}{0.400pt}}
\put(647.0,391.0){\rule[-0.200pt]{0.400pt}{4.577pt}}
\put(637.0,391.0){\rule[-0.200pt]{4.818pt}{0.400pt}}
\put(637.0,410.0){\rule[-0.200pt]{4.818pt}{0.400pt}}
\put(701.0,408.0){\rule[-0.200pt]{0.400pt}{4.818pt}}
\put(691.0,408.0){\rule[-0.200pt]{4.818pt}{0.400pt}}
\put(691.0,428.0){\rule[-0.200pt]{4.818pt}{0.400pt}}
\put(756.0,439.0){\rule[-0.200pt]{0.400pt}{5.300pt}}
\put(746.0,439.0){\rule[-0.200pt]{4.818pt}{0.400pt}}
\put(746.0,461.0){\rule[-0.200pt]{4.818pt}{0.400pt}}
\put(810.0,467.0){\rule[-0.200pt]{0.400pt}{5.782pt}}
\put(800.0,467.0){\rule[-0.200pt]{4.818pt}{0.400pt}}
\put(800.0,491.0){\rule[-0.200pt]{4.818pt}{0.400pt}}
\put(864.0,493.0){\rule[-0.200pt]{0.400pt}{6.986pt}}
\put(854.0,493.0){\rule[-0.200pt]{4.818pt}{0.400pt}}
\put(854.0,522.0){\rule[-0.200pt]{4.818pt}{0.400pt}}
\put(918.0,525.0){\rule[-0.200pt]{0.400pt}{6.745pt}}
\put(908.0,525.0){\rule[-0.200pt]{4.818pt}{0.400pt}}
\put(908.0,553.0){\rule[-0.200pt]{4.818pt}{0.400pt}}
\put(972.0,578.0){\rule[-0.200pt]{0.400pt}{7.468pt}}
\put(962.0,578.0){\rule[-0.200pt]{4.818pt}{0.400pt}}
\put(962.0,609.0){\rule[-0.200pt]{4.818pt}{0.400pt}}
\put(1026.0,576.0){\rule[-0.200pt]{0.400pt}{7.950pt}}
\put(1016.0,576.0){\rule[-0.200pt]{4.818pt}{0.400pt}}
\put(1016.0,609.0){\rule[-0.200pt]{4.818pt}{0.400pt}}
\put(1080.0,612.0){\rule[-0.200pt]{0.400pt}{7.468pt}}
\put(1070.0,612.0){\rule[-0.200pt]{4.818pt}{0.400pt}}
\put(1070.0,643.0){\rule[-0.200pt]{4.818pt}{0.400pt}}
\put(1134.0,622.0){\rule[-0.200pt]{0.400pt}{7.950pt}}
\put(1124.0,622.0){\rule[-0.200pt]{4.818pt}{0.400pt}}
\put(1124.0,655.0){\rule[-0.200pt]{4.818pt}{0.400pt}}
\put(1188.0,676.0){\rule[-0.200pt]{0.400pt}{8.913pt}}
\put(1178.0,676.0){\rule[-0.200pt]{4.818pt}{0.400pt}}
\put(1178.0,713.0){\rule[-0.200pt]{4.818pt}{0.400pt}}
\put(1242.0,702.0){\rule[-0.200pt]{0.400pt}{9.636pt}}
\put(1232.0,702.0){\rule[-0.200pt]{4.818pt}{0.400pt}}
\put(1232.0,742.0){\rule[-0.200pt]{4.818pt}{0.400pt}}
\put(1296.0,732.0){\rule[-0.200pt]{0.400pt}{9.636pt}}
\put(1286.0,732.0){\rule[-0.200pt]{4.818pt}{0.400pt}}
\put(1286.0,772.0){\rule[-0.200pt]{4.818pt}{0.400pt}}
\put(1350.0,778.0){\rule[-0.200pt]{0.400pt}{10.840pt}}
\put(1340.0,778.0){\rule[-0.200pt]{4.818pt}{0.400pt}}
\put(1340.0,823.0){\rule[-0.200pt]{4.818pt}{0.400pt}}
\put(1404.0,800.0){\rule[-0.200pt]{0.400pt}{10.359pt}}
\put(1394.0,800.0){\rule[-0.200pt]{4.818pt}{0.400pt}}
\put(161,303){\raisebox{-.8pt}{\makebox(0,0){$\Box$}}}
\put(215,295){\raisebox{-.8pt}{\makebox(0,0){$\Box$}}}
\put(269,291){\raisebox{-.8pt}{\makebox(0,0){$\Box$}}}
\put(323,285){\raisebox{-.8pt}{\makebox(0,0){$\Box$}}}
\put(377,296){\raisebox{-.8pt}{\makebox(0,0){$\Box$}}}
\put(431,311){\raisebox{-.8pt}{\makebox(0,0){$\Box$}}}
\put(485,328){\raisebox{-.8pt}{\makebox(0,0){$\Box$}}}
\put(539,344){\raisebox{-.8pt}{\makebox(0,0){$\Box$}}}
\put(593,365){\raisebox{-.8pt}{\makebox(0,0){$\Box$}}}
\put(647,401){\raisebox{-.8pt}{\makebox(0,0){$\Box$}}}
\put(701,418){\raisebox{-.8pt}{\makebox(0,0){$\Box$}}}
\put(756,450){\raisebox{-.8pt}{\makebox(0,0){$\Box$}}}
\put(810,479){\raisebox{-.8pt}{\makebox(0,0){$\Box$}}}
\put(864,507){\raisebox{-.8pt}{\makebox(0,0){$\Box$}}}
\put(918,539){\raisebox{-.8pt}{\makebox(0,0){$\Box$}}}
\put(972,594){\raisebox{-.8pt}{\makebox(0,0){$\Box$}}}
\put(1026,593){\raisebox{-.8pt}{\makebox(0,0){$\Box$}}}
\put(1080,627){\raisebox{-.8pt}{\makebox(0,0){$\Box$}}}
\put(1134,639){\raisebox{-.8pt}{\makebox(0,0){$\Box$}}}
\put(1188,694){\raisebox{-.8pt}{\makebox(0,0){$\Box$}}}
\put(1242,722){\raisebox{-.8pt}{\makebox(0,0){$\Box$}}}
\put(1296,752){\raisebox{-.8pt}{\makebox(0,0){$\Box$}}}
\put(1350,801){\raisebox{-.8pt}{\makebox(0,0){$\Box$}}}
\put(1404,821){\raisebox{-.8pt}{\makebox(0,0){$\Box$}}}
\put(1394.0,843.0){\rule[-0.200pt]{4.818pt}{0.400pt}}
\sbox{\plotpoint}{\rule[-0.500pt]{1.000pt}{1.000pt}}%
\put(161,123){\usebox{\plotpoint}}
\put(161.00,123.00){\usebox{\plotpoint}}
\put(179.10,133.14){\usebox{\plotpoint}}
\put(197.11,143.44){\usebox{\plotpoint}}
\put(214.94,154.05){\usebox{\plotpoint}}
\put(232.82,164.55){\usebox{\plotpoint}}
\put(250.80,174.89){\usebox{\plotpoint}}
\put(268.63,185.49){\usebox{\plotpoint}}
\put(286.90,195.33){\usebox{\plotpoint}}
\put(304.74,205.94){\usebox{\plotpoint}}
\put(322.78,216.18){\usebox{\plotpoint}}
\put(340.85,226.38){\usebox{\plotpoint}}
\put(358.68,236.98){\usebox{\plotpoint}}
\put(376.47,247.64){\usebox{\plotpoint}}
\put(394.53,257.83){\usebox{\plotpoint}}
\put(412.59,268.05){\usebox{\plotpoint}}
\put(430.64,278.27){\usebox{\plotpoint}}
\put(448.48,288.87){\usebox{\plotpoint}}
\put(466.47,299.21){\usebox{\plotpoint}}
\put(484.57,309.35){\usebox{\plotpoint}}
\put(502.42,319.92){\usebox{\plotpoint}}
\put(520.49,330.12){\usebox{\plotpoint}}
\put(538.30,340.78){\usebox{\plotpoint}}
\put(556.29,351.10){\usebox{\plotpoint}}
\put(574.40,361.24){\usebox{\plotpoint}}
\put(592.24,371.82){\usebox{\plotpoint}}
\put(610.18,382.26){\usebox{\plotpoint}}
\put(628.35,392.27){\usebox{\plotpoint}}
\put(646.38,402.54){\usebox{\plotpoint}}
\put(664.22,413.14){\usebox{\plotpoint}}
\put(682.07,423.73){\usebox{\plotpoint}}
\put(700.00,434.16){\usebox{\plotpoint}}
\put(718.10,444.30){\usebox{\plotpoint}}
\put(736.01,454.78){\usebox{\plotpoint}}
\put(754.29,464.62){\usebox{\plotpoint}}
\put(772.12,475.22){\usebox{\plotpoint}}
\put(790.09,485.59){\usebox{\plotpoint}}
\put(807.93,496.19){\usebox{\plotpoint}}
\put(825.83,506.68){\usebox{\plotpoint}}
\put(843.71,517.21){\usebox{\plotpoint}}
\put(861.94,527.12){\usebox{\plotpoint}}
\put(879.92,537.49){\usebox{\plotpoint}}
\put(898.02,547.63){\usebox{\plotpoint}}
\put(915.89,558.17){\usebox{\plotpoint}}
\put(933.80,568.64){\usebox{\plotpoint}}
\put(951.64,579.24){\usebox{\plotpoint}}
\put(969.60,589.63){\usebox{\plotpoint}}
\put(987.84,599.52){\usebox{\plotpoint}}
\put(1005.71,610.07){\usebox{\plotpoint}}
\put(1023.62,620.54){\usebox{\plotpoint}}
\put(1041.73,630.68){\usebox{\plotpoint}}
\put(1059.65,641.12){\usebox{\plotpoint}}
\put(1077.50,651.69){\usebox{\plotpoint}}
\put(1095.53,661.98){\usebox{\plotpoint}}
\put(1113.45,672.43){\usebox{\plotpoint}}
\put(1131.55,682.57){\usebox{\plotpoint}}
\put(1149.47,693.02){\usebox{\plotpoint}}
\put(1167.33,703.59){\usebox{\plotpoint}}
\put(1185.43,713.73){\usebox{\plotpoint}}
\put(1203.42,724.07){\usebox{\plotpoint}}
\put(1221.60,734.07){\usebox{\plotpoint}}
\put(1239.27,744.91){\usebox{\plotpoint}}
\put(1257.13,755.47){\usebox{\plotpoint}}
\put(1275.24,765.61){\usebox{\plotpoint}}
\put(1293.21,775.96){\usebox{\plotpoint}}
\put(1311.05,786.56){\usebox{\plotpoint}}
\put(1329.32,796.40){\usebox{\plotpoint}}
\put(1347.22,806.90){\usebox{\plotpoint}}
\put(1365.24,817.16){\usebox{\plotpoint}}
\put(1383.01,827.85){\usebox{\plotpoint}}
\put(1400.84,838.45){\usebox{\plotpoint}}
\put(1418.92,848.64){\usebox{\plotpoint}}
\put(1436.95,858.90){\usebox{\plotpoint}}
\put(1439,860){\usebox{\plotpoint}}
\end{picture}